\documentclass[lettersize,journal]{IEEEtran}
\usepackage{amsmath,amsfonts}

\usepackage{array}
\usepackage[caption=false,font=normalsize,labelfont=sf,textfont=sf]{subfig}
\usepackage{textcomp}
\usepackage{stfloats}
\usepackage{url}
\usepackage{makecell}
\usepackage{algorithm}
\usepackage{algpseudocode}
\usepackage{graphicx}
\usepackage{amsmath}
\usepackage{multirow}
\usepackage{subfig}
\usepackage{caption}
\usepackage{booktabs}
\usepackage{graphicx}
\usepackage{enumitem}
\usepackage{tcolorbox}
\usepackage{verbatim}
\usepackage{graphicx}
\usepackage{cite}
\begin{document}

\title{MirGuard: Towards a Robust Provenance-based Intrusion Detection System Against Graph Manipulation Attacks}

\author{Anyuan Sang, Lu Zhou, Li Yang, Junbo Jia, Huipeng Yang, Pengbin Feng and Jianfeng Ma

\thanks{Anyuan Sang, Junbo Jia, Huipeng Yang, Lu Zhou, and Li Yang are with the School of Computer Science and Technology, Xidian University, Xi'an, China.}
\thanks{Pengbin Feng and Jianfeng Ma are with the School of Cyber Engineering, Xidian University, Xi'an, China.}
\thanks{This work was supported in part by the National Key R\&D Program of China (2023YFB3106900), and the National Natural Science Foundation of China (grants No.62302362, 62402364, 62472337), in part by the Fundamental Research Funds for the Central Universities, and in part by the Innovation Fund of Xidian University under Grant YJSJ25012.}
\thanks{*Li Yang is the corresponding author. E-mail: yangli@xidian.edu.cn.}
}

\markboth{Journal of \LaTeX\ Class Files,~Vol.~14, No.~8, August~2021}%
{Shell \MakeLowercase{\textit{et al.}}: A Sample Article Using IEEEtran.cls for IEEE Journals}


\maketitle

\begin{abstract}
Learning-based Provenance-based Intrusion Detection Systems (PIDSes) have become essential tools for anomaly detection in host systems due to their ability to capture rich contextual and structural information, as well as their potential to detect unknown attacks. However, recent studies have shown that these systems are vulnerable to graph manipulation attacks, where attackers manipulate the graph structure to evade detection. While some previous approaches have discussed this type of attack, none have fully addressed it with a robust detection solution, limiting the practical applicability of PIDSes.

To address this challenge, we propose MirGuard, a robust anomaly detection framework that combines logic-aware multi-view augmentation with contrastive representation learning. Rather than applying arbitrary structural perturbations, MirGuard introduces Logic-Aware Noise Injection (LNI) to generate semantically valid graph views, ensuring that all augmentations preserve the underlying causal semantics of the provenance data. These views are then used in a Logic-Preserving Contrastive Learning framework, which encourages the model to learn representations that are invariant to benign transformations but sensitive to adversarial inconsistencies. Comprehensive evaluations on multiple provenance datasets demonstrate that MirGuard significantly outperforms state-of-the-art detectors in robustness against various graph manipulation attacks without sacrificing detection performance and efficiency. Our work represents the first targeted study to enhance PIDS against such adversarial threats, providing a robust and effective solution to modern cybersecurity challenges.
\end{abstract}

\begin{IEEEkeywords}
Intrusion Detection System, Provenance Graph, Graph Manipulation Attack.
\end{IEEEkeywords}

\section{Introduction}
\IEEEPARstart{A}{dvanced} Persistent Threats (APTs) have become increasingly prevalent, posing significant risks to global cybersecurity \cite{Aptnotes}. These sophisticated and stealthy attacks target critical infrastructure, government systems, and private enterprises, often leading to severe data breaches, financial losses, and national security threats. The persistent nature of APTs allows attackers to maintain a foothold within compromised networks for extended periods, enabling them to exfiltrate sensitive information and disrupt operations, causing widespread harm to society and the economy.

Provenance graphs, which capture the causal relationships between system entities and events, have become a valuable foundation for behavior-based intrusion detection. These graphs provide rich contextual information that enables detailed analysis of system activity and potential attack chains \cite{inam2022sok,dong2023we}. Detection methods based on provenance graphs can be broadly categorized into two approaches: knowledge-based \cite{hossain2017sleuth,milajerdi2019holmes,hassan2020tactical,zhu2023aptshield} and learning-based techniques \cite{yang2023prographer,wang2020you,wang2022threatrace,zengy2022shadewatcher,jia2024magic,cheng2024kairos,rehman2024flash}. Knowledge-based methods rely on predefined rules or metrics to perform anomaly detection within the graph. However, their dependence on prior knowledge and inability to capture advanced, deep features have driven researchers toward learning-based approaches.

Learning-based detection methods leverage various levels of graph embedding techniques in upstream tasks (graph learning), such as node embedding \cite{wang2022threatrace,jia2024magic}, edge embedding \cite{wang2020you,zengy2022shadewatcher}, and subgraph embedding \cite{yang2023prographer,shen2019attack2vec}, to derive expressive representations of the provenance graph. These techniques effectively capture both contextual and structural information. Subsequently, downstream detection algorithms, including outlier detection and vector similarity analysis, are applied to identify anomalies and detect potential attacks.

Although existing methods have demonstrated effective detection performance, recent graph manipulation attack strategies pose significant challenges to these detectors \cite{goyal2023sometimes, Kunal2023Evading, sang2024obfuscating}. graph manipulation attacks involve attackers forging interaction information of malicious processes, such as adding sufficient edges connecting to benign nodes, to shift their representation in the embedding space and evade detection. This vulnerability arises from an inherent limitation of machine learning models, where small perturbations can lead to high-confidence misclassification \cite{croce2020reliable,athalye2018obfuscated}. To the best of our knowledge, only a few studies \cite{han2021sigl,wang2022threatrace,jia2024magic,rehman2024flash} have briefly explored the impact of such attacks on provenance graph-based detection methods, and even fewer have proposed targeted mitigation strategies. Generic robustness enhancement strategies, such as adversarial training \cite{wang2017adversary,aldujaili2018adversarial}, face practical challenges in the PIDS domain due to the scarcity of malicious samples and may be ineffective against potential unseen attacks. Therefore, current defense mechanisms are inadequate for countering mimicry attacks, highlighting an urgent need for a novel robustness enhancement approach that improves the intrinsic robustness of detection models.

In this paper, we propose \textbf{MirGuard}, a novel anomaly detection method based on provenance graphs, designed to enhance robustness against graph manipulation attacks while maintaining high detection accuracy. Provenance graphs are vulnerable to such attacks, where adversaries mimic benign behaviors to conceal malicious activities and evade detection \cite{goyal2023sometimes, Kunal2023Evading, sang2024obfuscating}. In this work, we analyze the typical workflow of PIDS and identify two primary types of graph manipulation attacks (as detailed in the threat model section): graph poisoning attacks during the training phase and graph pollution attacks during the detection phase. To counter such attacks, MirGuard leverages a multi-view learning strategy that combines structured graph augmentation with contrastive learning. The key idea is to force the model to learn representations that are invariant to local perturbations and sensitive to global malicious patterns. 

Specifically, MirGuard applies a logic-aware graph augmentation strategy, which ensures that all perturbations conform to structural semantics defined by the provenance context (e.g., disallowing file-to-network or network-to-network edges). This results in more realistic adversarial simulations compared to random augmentations. These augmentations disrupt attacker-crafted patterns and encourage the model to focus on more stable, graph-level semantics.

Based on these augmented views, MirGuard employs a contrastive learning framework that encourages semantic consistency across views while distinguishing unrelated behaviors. Unlike conventional approaches such as GraphCL \cite{you2020graphCL} or MVGRL \cite{hassani2020MVGRL}, our method emphasizes semantic consistency rooted in domain-specific logic, rather than superficial structural similarity alone. By learning representations invariant to benign-appearing manipulations but sensitive to semantic inconsistencies, MirGuard achieves strong robustness against both poisoning and evasion attacks. Our evaluations demonstrate that this design leads to improved generalization and more reliable anomaly detection in complex and adversarial environments.

After obtaining robust graph representations, MirGuard employs an unsupervised anomaly detection mechanism based on KMeans clustering. In the training phase, KMeans is used to partition the embedding space into \(k\) clusters. The centroids of these clusters, along with the average intra-cluster distance across training samples, are retained as references. During inference, each test sample is evaluated by computing its Euclidean distances to all cluster centroids. The minimum distance is taken as the initial anomaly score, which is then normalized by the global average distance. If the normalized score exceeds a predefined threshold, the sample is flagged as anomalous. This centroid-based detection strategy enables MirGuard to perform efficient and scalable anomaly detection, significantly reducing inference overhead while preserving high detection performance.

To comprehensively evaluate the efficiency of MirGuard, we utilized widely adopted provenance datasets, including DARPA TC THEIA, CADETS, TRACE \cite{TC3}, the Streamspot dataset \cite{Streamspot_data}, and the Unicorn Wget dataset \cite{han2020unicorn}. We also employed several state-of-the-art graph learning-based anomaly detectors, such as Threatrace \cite{wang2022threatrace}, MAGIC \cite{jia2024magic}, and FLASH \cite{rehman2024flash}, as baselines. To thoroughly assess MirGuard's robustness against graph manipulation attacks, we implemented five types of such attacks during both the detection and training phases, based on prior evasion studies \cite{goyal2023sometimes,Kunal2023Evading,sang2024obfuscating}. In our experiments, we first evaluated MirGuard's resistance to different attack types and compared its robustness with that of current state-of-the-art detection schemes. We then discussed whether MirGuard sacrifices detection performance to achieve robustness. Next, we demonstrated the rationale and necessity of MirGuard's module design through ablation experiments. Finally, we evaluated the overhead of MirGuard and discussed the impact of different parameter settings on its performance.

Our contributions are summarized as follows:
\begin{itemize}
    \item To the best of our knowledge, we are the first to specifically enhance the robustness of PIDS models against graph manipulation attacks.
    \item We propose a novel graph learning-based PIDS, MirGuard, which is designed with a unique multi-view augmentation strategy and employs a contrastive learning mechanism to train the model. This enables MirGuard to achieve strong robustness against graph manipulation attacks while maintaining detection performance comparable to state-of-the-art detectors.
    \item We implemented five types of attacks across both the training and detection phases and conducted comprehensive evaluations of MirGuard's robustness and detection performance on multiple datasets. Experimental results demonstrate that, compared to baseline systems, MirGuard exhibits exceptional robustness against graph manipulation attacks without compromising detection performance or incurring additional detection overhead (achieving an average F1-score of over 96\% with less than 10\% AUC drop under graph manipulation attacks).
    % \item To further support the research community, we have open-sourced the implementation of MirGuard. \footnote{ MirGuard is available on \url{}}
\end{itemize}

\section{Background}
\subsection{Graph Manipulation Attacks}
Graph manipulation attacks \cite{goyal2023sometimes, Kunal2023Evading, sang2024obfuscating} pose a significant challenge to graph-based systems by strategically altering graph structures to evade detection or degrade model performance. These attacks often target critical graph elements, such as nodes, edges, or features, to disguise malicious behavior as benign or disrupt the learning process of graph-based models. For instance, attackers may inject fake nodes or edges to obscure critical relationships or modify existing features to mimic benign entities, making it harder to detect anomalies. In the context of provenance graphs, these attacks exploit the graph’s structural and semantic complexity, where malicious subgraphs are embedded within larger benign structures, allowing adversaries to manipulate local patterns while preserving global consistency. This obfuscation enables attackers to bypass anomaly detection methods that heavily rely on local or static patterns. Addressing such attacks requires robust graph-based methods that can capture invariant global features and distinguish subtle manipulations, ensuring resilience against adversarial perturbations.

\subsection{Provenance-based IDS}
Since the provenance graph can express the relationship between system operating entities in time, existing research has used this feature to build an IDS based on the provenance graph. Including detection schemes based on knowledge labels \cite{hossain2017sleuth,milajerdi2019holmes,hassan2020tactical,zhu2023aptshield}, these schemes construct a series of matching rules based on expert knowledge to match in the origin graph to detect anomalies. Based on the statistics IDS scheme \cite{hassan2019nodoze,fang2022back,liu2018towards}, they use the structural feature information of the graph, including: abnormality, discrepancy, time correlation and other features to analyze in the graph to detect anomalies. Recently, more learning-based IDS solutions have been proposed \cite{alsaheel2021atlas,zengy2022shadewatcher,han2021sigl,yang2023prographer,jia2024magic,shen2019attack2vec}. These solutions use models such as graph representation learning and sequence learning to extract high-dimensional features from graphs to perform anomaly detection in downstream tasks.

\begin{figure}[t]
    \centering
    
    \includegraphics[width=1\linewidth]{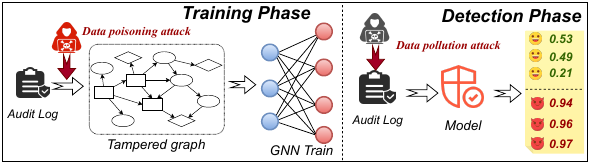}
    \caption{The classic detection processes of PIDS identified two types of graph manipulation attacks: data poisoning attacks during the training phase and data pollution attacks during the detection phase.}
    \label{fig: adv_analysis}
    
\end{figure}

\section{Motivation Example} 

\begin{figure}[t]
    \centering
    \includegraphics[width=1\linewidth]{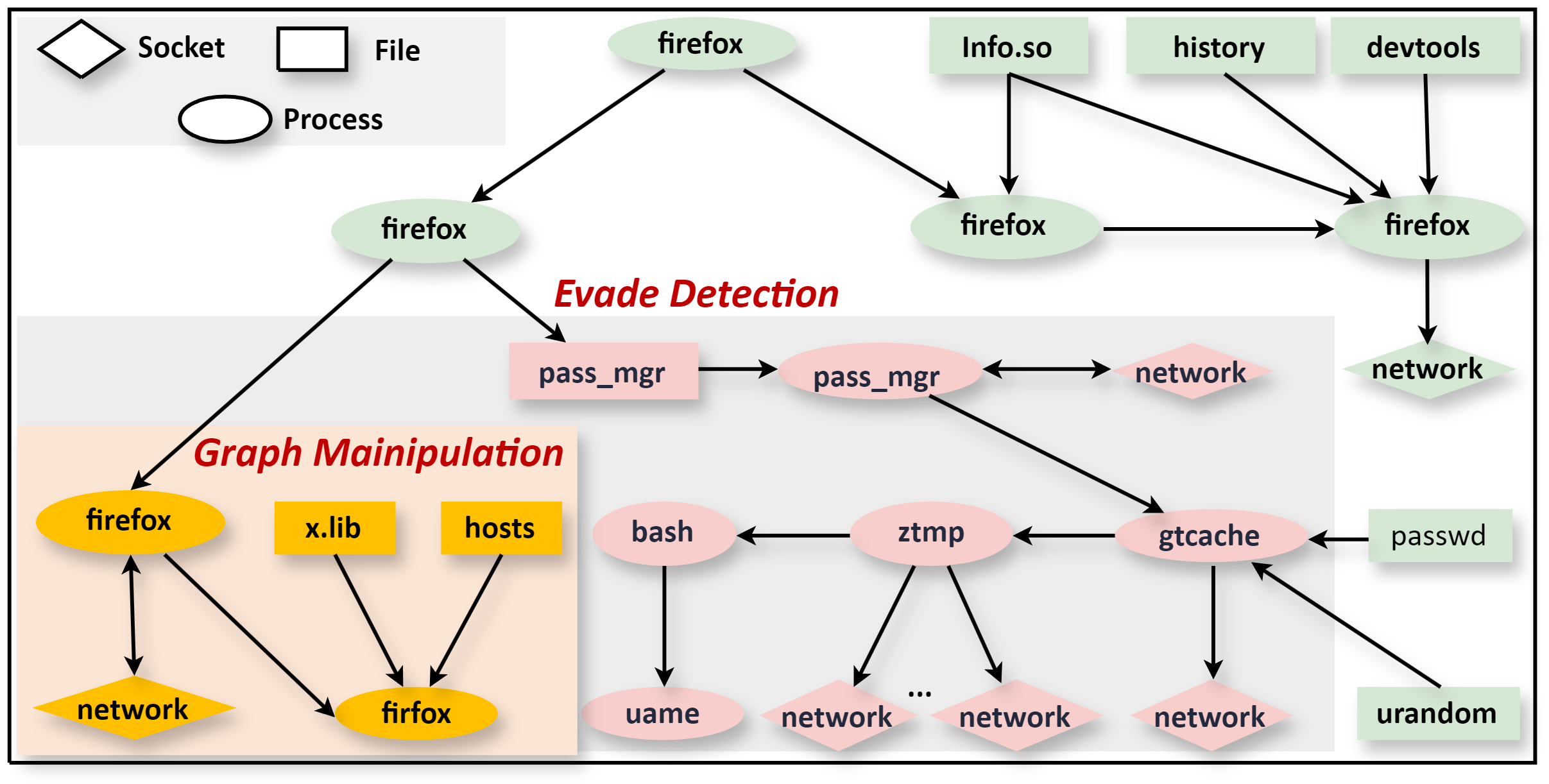}
    \caption{In the provenance graph of the TC E3 browser extension attack, we considered a graph manipulation attack proposed by \cite{goyal2023sometimes}, where attackers could manipulate the graph structure by inserting benign subgraphs into the attack subgraph to evade detection. The green nodes represent benign nodes engaged in normal activities, red nodes represent attack nodes, and yellow nodes indicate attack nodes added by the attacker.}
    \label{fig: motivation-example}
\end{figure}

This scenario illustrates an APT attack conducted through a browser extension in the DARPA TC E3 dataset. Figure \ref{fig: motivation-example} provides a simplified visualization of this attack. The attack was initiated when the victim visited a malicious website that exploited a vulnerability in the \textit{pass\_mgr} extension of the Firefox browser. The attacker leveraged this vulnerability to download a program named \textit{gtcache}. The \textit{gtcache} program connected with the attacker and executed data theft operations. Additionally, it installed another program, \textit{ztmp}, to gather system configuration details and perform port scans on the target network for internal reconnaissance. Notably, in this attack scenario, we introduce a manipulation strategy where the attacker alters the graph structure by inserting benign subgraphs into the attack subgraph to evade detection.

This attack poses a significant challenge to existing learning-based detection methods, especially those relying on graph embedding techniques such as GraphSAGE, GNNs, and Graph2Vec \cite{wang2022threatrace, yang2023prographer,zengy2022shadewatcher}. These methods are susceptible to graph manipulation attacks, which can modify the neighborhood structure of malicious nodes, causing them to resemble benign nodes more closely. As a result, the embeddings learned during training may increasingly resemble benign behavior, significantly impairing the model’s ability to differentiate between malicious and benign activities. This phenomenon, known as evasion attacks, occurs when the attacker manipulates the graph such that malicious nodes are embedded in regions of the graph space typically occupied by benign nodes. Consequently, the detection model, trained on these altered embeddings, becomes more prone to evasion, resulting in a decline in its overall robustness. These challenges underscore the need for detection systems that can not only learn effective representations of graph data but also remain robust against adversarial manipulations designed to conceal malicious behaviors.

\begin{figure*}[htbp]
    \centering
    \includegraphics[width=1\linewidth]{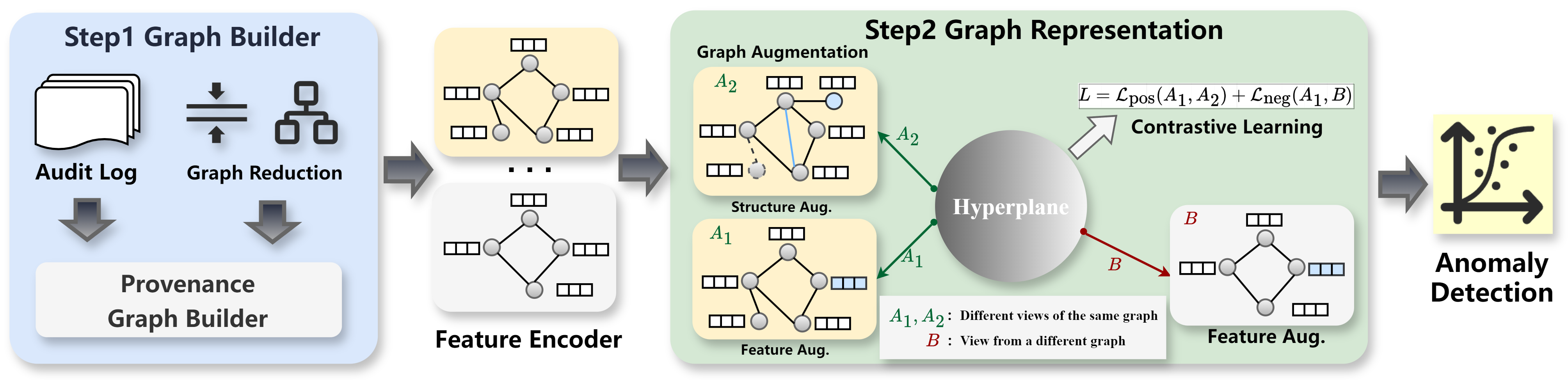}
    \caption{Overview of MirGuard's architecture.}
    \label{fig: overview}
\end{figure*}

\section{Threat Model \& Assumptions}
Our experimental environment relies on a Trusted Computing Base (TCB) consisting of the operating system, auditing framework, and provenance analysis tools. We assume that all components within the TCB function correctly throughout the entire process from installation to execution. This assumption is standard in existing provenance-based detectors. Hardware trojans and side-channel attacks that cannot be captured through audit mechanisms are not considered in this paper. In addition, we assume that the integrity of the output audit data is guaranteed by existing secure provenance and integrity audit systems \cite{bates2015trustworthy, pasquier2017practical, paccagnella2020custos, zeng2022palantir, zhang2024hitchhiker}.

In our robustness evaluation experiments, we analyzed the typical processing pipeline of provenance-based systems, as shown in Figure~\ref{fig: adv_analysis}, and identified two types of graph manipulation attacks: data poisoning attacks during the model training phase and data pollution attacks during the detection phase. Previous studies \cite{goyal2023sometimes, Kunal2023Evading, sang2024obfuscating} assumed that adversaries can manipulate the structure of the provenance graph to launch attacks against the detector. This falls into the category of data pollution attacks. Based on this, we extend the attack model by introducing a stronger assumption in which adversaries can also inject crafted graph perturbations into the audit logs during the training phase. This results in data poisoning attacks that affect the model’s training outcomes.

\section{Design}

As shown in Fig.\ref{fig: overview} MirGuard comprises three main components: (1) Graph Builder, (2) Graph Representation, and (3) Anomaly Detection.

In the graph builder module, MirGuard processes system audit logs to construct the provenance graph, where nodes represent system entities and edges denote interactions between them. Edge compression techniques are employed to merge redundant nodes and edges, optimizing the graph structure and reducing computational complexity. Additionally, batch-based provenance graph construction is implemented to handle large-scale data by splitting the graph into smaller batches. Node and edge types are extracted for subsequent feature encoding in the representation module.

The core of MirGuard lies in the graph representation module, which includes feature encoding, graph augmentation, and contrastive learning. First, node and edge features are encoded using one-hot encoding to standardize the input. Then, GNN, such as a Graph Attention Network (GAT), is employed to extract higher-order structural and semantic features, capturing both local and global dependencies. Graph augmentation introduces controlled perturbations to simulate adversarial scenarios, enhancing the model's ability to learn invariant representations. Finally, a contrastive learning framework aligns embeddings of augmented views while maintaining separation between distinct graphs, ensuring robustness against adversarial manipulations.

The anomaly detection module employs a KMeans-based detection method to identify anomalous nodes in the graph. Although various classifiers were considered, KMeans-based detection demonstrated superior performance in our evaluations, as detailed in Section \ref{evl: ablation}.

\subsection{Graph Builder} \label{sec:5.1}
Our system accepts streaming system audit logs and constructs the provenance graph, similar to previous research \cite{king2003backtracking,king2005enriching}. It consists of three main components. First, MirGuard streams and extracts audit logs in batches from existing operating systems, such as Windows ETW logs or Linux audit logs. These logs contain information about interactions between system entities, including files, processes, and networks. Next, MirGuard extracts and processes this log information. Specifically, for each audit log within a batch, it extracts the fundamental components representing the nodes and edges of the provenance graph: the quadruple \textit{(src, dst, timestamp, edge type)}, where \textit{src} denotes the process node, \textit{dst} represents the file or network node, \textit{timestamp} indicates the time when the event occurred, and \textit{edge type} specifies the type of edge. Finally, to accelerate model training and reduce computational complexity, we adopt multi-class graph denoising techniques from prior studies, removing only redundant nodes and those irrelevant to attack detection. MirGuard utilizes the CPR (Causal Persistent Reduction) method \cite{xu2016high} for edge processing, retaining only one instance of edges that appear multiple times between two nodes within a short time window. Additionally, during graph construction, orphaned nodes and faulty nodes (potentially generated by logging errors) that are unrelated to the attack investigation are removed.

\subsection{Graph Representation}
Graph representation in MirGuard involves a systematic pipeline to transform the raw provenance graph into robust embeddings suitable for anomaly detection. This process begins with feature encoding, where node and edge attributes are represented using one-hot encoding and refined through GNNs to capture both local and global structural dependencies. Following this, graph augmentation introduces controlled perturbations to simulate adversarial scenarios, enhancing the model's ability to learn invariant representations. Finally, a contrastive learning framework aligns embeddings of augmented views while maintaining separation between distinct graphs, ensuring robustness against adversarial manipulations and capturing meaningful graph semantics. Together, these steps enable MirGuard to construct high-quality graph embeddings that are both expressive and resilient, forming the foundation for reliable detection in complex environments.

\subsubsection{Feature Encoding}
MirGuard begins by encoding the raw provenance graph’s node and edge attributes using one-hot encoding. Each node and edge is represented by its type, and one-hot encoding is applied to generate a categorical feature vector. This process transforms discrete attributes, such as node types (e.g., processes, files) and edge types (e.g., read, write), into binary vectors that preserve their distinct semantics.

Once the one-hot encoding is complete, MirGuard employs a Graph Neural Network (GNN), such as a Graph Attention Network (GAT), to extract higher-order structural and semantic features. The GNN processes the graph by aggregating information from neighboring nodes and edges, capturing both local dependencies and global contextual patterns. For a node \(v\), its feature representation \(h_v^{(l+1)}\) at layer \(l+1\) is computed as:
\[
h_v^{(l+1)} = \sigma \left( W^{(l)} \cdot \text{AGG}\left(\{h_u^{(l)} \mid u \in \mathcal{N}(v)\}\right) + b^{(l)} \right),
\]
where \(h_v^{(l)}\) is the feature vector of node \(v\) at layer \(l\), \(\mathcal{N}(v)\) represents the neighbors of \(v\), \(\text{AGG}(\cdot)\) is an aggregation function (e.g., sum or mean), \(W^{(l)}\) and \(b^{(l)}\) are learnable parameters, and \(\sigma\) is an activation function (e.g., ReLU). This encoding process generates a dense feature vector for each node and edge, capturing both their individual properties and relational information.

The output of the GNN serves as the input for subsequent graph augmentation and contrastive learning steps.

\subsubsection{Logic-Aware Graph Augmentation}

MirGuard employs logic-aware graph augmentation strategies to simulate potential adversarial attacks while maintaining the semantic plausibility of the provenance graph. These augmentations include edge augmentation (EA), node augmentation (NA), and feature augmentation (FA), each applied to generate perturbed views of the original graph. The intensity of each augmentation operation is controlled by the hyperparameter \(\gamma\), which specifies the proportion of nodes or edges to be modified.

To ensure the realism and logical validity of augmented graphs, we implement a strategy called \textbf{Logic-Aware Noise Injection (LNI)} as shown in Table \ref{tab:logic_augmentation}. This strategy enforces rationality constraints during augmentation to prevent the generation of semantically invalid structures. For instance, in edge augmentation, we prohibit the addition of edges that directly connect two network nodes or connect a file node to a network node—such configurations violate causal semantics in provenance graphs.

\begin{figure}[t]
    \centering
    \includegraphics[width=1\linewidth]{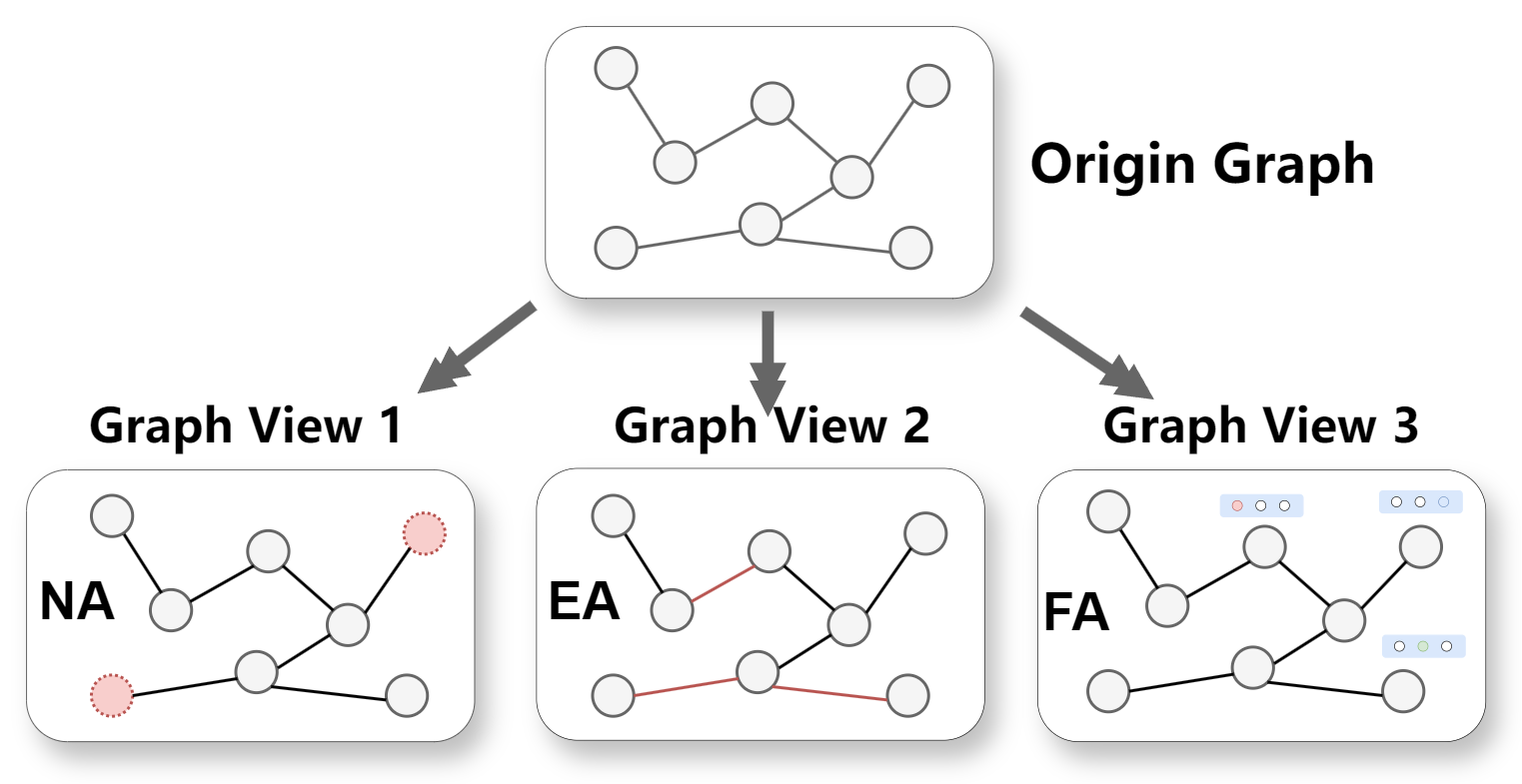}
    \caption{Examples of logic-aware graph augmentation. Edge and node operations are constrained to preserve provenance semantics.}
    \label{fig: graph_aug}
\end{figure}

\begin{table}[h]
\centering
\caption{Logic-Aware Edge Augmentation Rules in Provenance Graphs}
\begin{tabular}{lccc}
\toprule
\textbf{Source} & \textbf{Destination} & \textbf{Allowed} & \textbf{Edge} \\
\midrule
Process              & File            &  \checkmark        & read/write \\
Process              & Network         &  \checkmark        & connect/send/recv  \\
File                 & Process         &   \checkmark       & exec/load \\
Process              & Process         &   \checkmark       & fork/clone \\
Network              & Process         & $\times$  & violates causality \\
File                 & Network         & $\times$  & no direct communication \\
Network              & File            & $\times$  & no direct communication \\
Network              & Network         & $\times$  & meaningless edge \\
\bottomrule
\end{tabular}
\label{tab:logic_augmentation}
\end{table}

\paragraph{Edge Augmentation (EA)}
Edge augmentation modifies the graph structure by adding or removing edges under logic constraints. The edge set \(E'\) is modified as follows:

\[
E' = E \cup \{(u, v)\},
\]
This operation adds an edge between nodes \(u\) and \(v\) only if it does not already exist \textit{and} the connection satisfies domain-specific logic rules.

\[
E' = E \setminus \{(u, v)\},
\]
Alternatively, an existing edge can be removed. The perturbation intensity is controlled by \(\gamma\); for example, 20\% of edges are randomly selected for addition or removal, subject to logical validity.

\paragraph{Node Augmentation (NA)}
Node augmentation involves adding or removing nodes along with their associated edges, while ensuring logical consistency in their insertion or removal context:

\[
V' = V \cup \{v'\},
\]
This means a new node \(v'\) is added to the graph and linked to others only through permissible edge types.

\[
V' = V \setminus \{v\}, \quad E' = E \setminus \{(u, v)\},
\]
An existing node \(v\) may be removed along with all its connected edges. Node augmentation is also governed by \(\gamma\); for example, 20\% of the nodes are selected for addition or deletion.

\paragraph{Feature Augmentation (FA)}
Feature augmentation modifies node attributes while preserving semantic alignment. The feature vector of node \(v\) is replaced with that of another node \(w\) of the same type:

\[
X'_v = X_w, \quad w \sim \{u \in V \mid \text{type}(u) = \text{type}(v)\},
\]
This simulates adversarial feature manipulation without disrupting node-type semantics.

Collectively, these logic-aware augmentations enhance the model’s ability to learn invariant patterns and detect adversarial perturbations that preserve surface semantics but violate causal consistency. They prepare the model for challenging attack scenarios while retaining the integrity of the graph's structural and semantic foundations.

\subsubsection{Logic-Preserving Contrastive Learning}

To further improve robustness, MirGuard introduces a contrastive learning framework tailored to provenance graphs, with an emphasis on preserving the underlying causal semantics. Unlike general-purpose contrastive learning frameworks such as GraphCL~\cite{you2020graphCL} or MVGRL \cite{hassani2020MVGRL}, which rely on random augmentations and structure-based similarity, our approach incorporates domain-aware augmentations and logic consistency.

Given two augmented views \( G_i \) and \( G_j \) of the same original graph \( G \), the encoder generates graph-level embeddings \( z_i \) and \( z_j \), which are passed through a two-layer projection head:

\[
p_v = \text{ReLU}(W_p^{(1)} z_v + b_p^{(1)}), \quad 
\hat{p}_v = W_p^{(2)} p_v + b_p^{(2)}.
\]

We then compute the contrastive loss:
\[
\mathcal{L} = -\log \frac{\exp(\text{sim}(\hat{p}_i, \hat{p}_j) / \tau)}{\sum_{k=1}^N \exp(\text{sim}(\hat{p}_i, \hat{p}_k) / \tau)},
\]
where \(\text{sim}(\cdot, \cdot)\) denotes cosine similarity, and \(\tau\) is a temperature parameter. Positive pairs \((\hat{p}_i, \hat{p}_j)\) originate from different views of the same graph that preserve logical structure, while negatives \(\hat{p}_k\) come from unrelated graphs.

This design ensures that the learned embeddings reflect consistent high-level behaviors rather than superficial structural features. It enables the model to resist manipulation that mimics graph topology while violating semantic logic, which is especially critical for provenance-based anomaly detection.

\subsection{Anomaly Detection}

To determine the most effective self-supervised anomaly detection mechanism, we evaluated several candidate classifiers, including Local Outlier Factor (LOF) \cite{breunig2000lof}, One-Class SVM \cite{scholkopf1999OC-SVM}, KMeans \cite{guo2003knn}, and Isolation Forest \cite{liu2008isolation}. As detailed in Section \ref{evl: ablation}, KMeans demonstrated superior performance in terms of detection accuracy. Consequently, MirGuard employs a KMeans-based anomaly detector, which includes training and detection phases. 

In our method, we apply K-means to partition the embedding space into \( k \) clusters and retain all cluster centroids for subsequent anomaly detection.

During the detection phase, each new embedding vector is evaluated by computing its distance to the nearest cluster centroid. The anomaly score \( S_i \) for a data point \( x_i \) is defined as the Euclidean distance to the closest centroid among the \( k \) clusters:

\[
S_i = \min_{j=1,\ldots,k} \left\| x_i - c_j \right\|,
\]

where \( c_j \) denotes the \( j \)-th cluster centroid obtained from training.

To ensure comparability across datasets and feature scales, we normalize the raw anomaly score using the mean nearest-centroid distance computed on the training set, denoted as \( D_{\text{mean}} \). This is defined as:

\[
D_{\text{mean}} = \frac{1}{N} \sum_{i=1}^{N} \min_{j=1,\ldots,k} \left\| x_i - c_j \right\|,
\]

where \( N \) is the number of training samples. The normalized anomaly score \( \tilde{S}_i \) is given by:

\[
\tilde{S}_i = \frac{S_i}{D_{\text{mean}}}
\]

A data point is considered anomalous if its normalized score exceeds a predefined threshold \( \theta \):

\[
\text{Anomaly}(x_i) = 
\begin{cases}
1 & \text{if } \tilde{S}_i > \theta, \\
0 & \text{otherwise}.
\end{cases}
\]

Given the large-scale nature of provenance data, this centroid-based evaluation strategy significantly reduces the inference overhead compared to pairwise distance-based approaches such as KNN \cite{jia2024magic}, while maintaining effective anomaly detection performance, especially for large-scale provenance data.

\section{Evluation}

In this section, we evaluate the performance of MirGuard by addressing the following research questions (RQs):

\noindent\begin{itemize}
    \item \textbf{RQ1:} How is MirGuard’s detection efficiency compared to baseline methods?
    \item \textbf{RQ2:} Does MirGuard successfully improve robustness against graph manipulation attacks compared to its baselines?
    \item \textbf{RQ3:} To what extent do the structured augmentations and multi-view contrastive learning contribute to MirGuard's ability to counteract graph manipulation attacks and detect malicious behaviors?
    \item \textbf{RQ4:} Does MirGuard introduce significant computational overhead compared to existing PIDSes?
\end{itemize}

\subsection{Experiments Setup} \label{eva: setup}
In data processing, we adopted the log transformers in MAGIC \cite{jia2024magic} for processing streaming audit logs, including StreamSpot \cite{manzoor2016streamspot}, Camflow \cite{pasquier2017practical} and DARPA TC Dataset \cite{TC3}. Networkx is used to construct the provenance graph. The graph indicates that the learning module is implemented by Pytorch \cite{paszke2019pytorch} and DGL \cite{wang2019dDGL}.

\noindent \textbf{Parameter settings.} For the setup of MirGuard, the learning rate \textit{lr} is set to 0.001. We use a 2-layer GAT encoder, and in data augmentation, the augmentation ratio is set to 0.5. The training batch size is 50 with \textit{d} set to 64 on the DARPA TC dataset.

\noindent \textbf{Datasets.} We evaluated the performance of MirGuard under three open-source datasets: DARPA Engagement TC E3 \cite{TC3}, Streamspot and, Unicorn Wget. All three datasets are inconsistent in the scenarios they target and the granularity of their detections, and thus we believe they are able to provide insights into the performance of the system. The detail of the dataset description are as follows:

\begin{table}[h]
  \centering
  \caption{Dataset Statistics for Streamspot and Unicorn Wget}
  \resizebox{0.48\textwidth}{!}{%
    \begin{tabular}{l|cccc}
      \hline
      \textbf{Dataset} & \textbf{Graph Pieces} & \textbf{Entities} & \textbf{Interactions} & \textbf{Size (GB)} \\
      \hline
      \multirow{6}{*}{Streamspot} 
        & \multirow{6}{*}{100} 
        & 8,292 & 113,229 & \multirow{6}{*}{2.8} \\
        &       & 8,636 & 112,958 &  \\
        &       & 8,989 & 294,903 &  \\
        &       & 8,830 & 310,814 &  \\
        &       & 6,826 &  37,382 &  \\
        &       & 8,890 &  28,423 &  \\
      \hline
      \multirow{2}{*}{Unicorn Wget} 
        & 125 & 265,424 & 975,226 & \multirow{2}{*}{76} \\
        & 25  & 257,156 & 949,887 &  \\
      \hline
    \end{tabular}%
  }
\end{table}

\begin{table}[h]
  \centering
  \caption{Dataset Statistics for DARPA E3 }
    \renewcommand{\arraystretch}{1.1} % 调整行距
  \resizebox{0.47\textwidth}{!}{%
    \begin{tabular}{@{}l|cccc@{}}
      \hline
      \textbf{Dataset} & \textbf{Benign Nodes} & \textbf{Abnormal Nodes} & \textbf{Edges} & \textbf{Size (GB)} \\
      \hline
      E3 Trace & 3,220,596 & 68,082 & 4,080,457 & \multirow{3}{*}{67} \\
      E3 Cadets & 1,614,189 & 12,846 & 3,303,264 & \\
      E3 Theia & 3,505,326 & 25,362 & 10,929,710 & \\
      \hline
    \end{tabular}%
  }
\end{table}

\begin{itemize}
    \item \textbf{DARPA TC dataset.} The DARPA TC dataset is a benchmark dataset provided by DARPA for evaluating cybersecurity and intrusion detection systems. It was collected from networks during adversarial engagements. The red team conducted APT attacks using various vulnerabilities to exfiltrate information. Our evaluation includes the TRACE, CADETS, and THEIA subdatasets, which contain millions of entities and interaction records. We used the ground truth information provided by the Threatrace \cite{wang2022threatrace} to perform entity-level detection and conduct attack investigations.
    \item \textbf{Unicorn Wget dataset.} The Wget dataset was designed by the authors of Unicorn \cite{han2020unicorn} to simulate attack scenarios. It uses the Camflow \cite{pasquier2017practical} system to collect 150 batches of audit logs, with 125 batches containing no attack processes and 25 batches containing supply chain attacks. These attacks are carefully crafted to mimic benign entity interactions, making this dataset challenging to identify due to its large data volume and stealthy attacks. We will perform graph-level detection on this dataset as in previous approaches.
    \item \textbf{StreamSpot dataset.} The StreamSpot dataset is a publicly available dataset provided by the authors of StreamSpot \cite{manzoor2016streamspot}, containing 600 information flow graphs. These graphs come from five benign scenarios and one attack scenario. Each scenario runs 100 times, generating 100 graphs using the Linux SystemTap Logging System. The five benign scenarios simulate normal user behavior, while the attack scenario simulates a drive-by download attack. We performed graph-level anomaly detection on the StreamSpot dataset, similar to previous studies \cite{han2020unicorn,wang2022threatrace}, as it only provides graph-level ground truth.

\end{itemize}

\begin{table}[t]
\centering
\caption{Details of Graph Manipulation Attacks}
\label{tab:graph_attacks_detail}
  \renewcommand{\arraystretch}{1.1} % 调整行距
\begin{tabular}{l|cccc}
\hline
\textbf{Phase} & \textbf{Attack Type} & \textbf{Target}                      & \textbf{Rate (\(y\))} \\ 
\hline
\multirow{3}{*}{Detection} 
    & GSPA     & Node      & \(y\)        \\ 
    & GFPA    & Edge          & \(y\)        \\ 
    & CGPA    & Node \& Edge      & \(0.5y + 0.5y\)   \\ 
\hline
\multirow{2}{*}{Training} 
    & SPA  & Node \& Edge  & \(0.5y + 0.5y\)       \\ 
    & FPA   & Node      & \(y\)             \\ 
\hline
\end{tabular}
\end{table}

\noindent \textbf{Baselines.} To comprehensively evaluate the detection performance of MirGuard, we compare it with state-of-the-art (SOTA) and open-source graph-based methods in the PIDS domain, including Threatrace \cite{wang2022threatrace}, MAGIC \cite{jia2024magic}, and FLASH \cite{rehman2024flash}. It is worth noting that several other approaches were not included in our comparison for the following reasons:

First, since MirGuard is a graph-based anomaly detection method, we excluded signature-based methods \cite{milajerdi2019holmes,zhu2023aptshield,wang2024CAPTION}, priority-based approaches \cite{fang2022back,hassan2019nodoze,liu2018towards}, and graph sketch-based techniques \cite{han2020unicorn}. Additionally, some recent works \cite{cheng2024kairos,jian2025ORTHRUS} adopt finer-grained root node labeling strategies, which differ significantly from our threat model and would hinder a fair comparison. As such, these methods were also excluded.

Second, as noted by the authors of \cite{cheng2024kairos}, many learning-based detectors in the PIDS domain, such as ProvDetector \cite{wang2020you}, ShadeWatcher \cite{zengy2022shadewatcher}, RCAID \cite{goyal2024rcaid}, and ProGrapher \cite{yang2023prographer}, are not fully open-source. Reproducing these methods solely based on their published descriptions may introduce experimental bias; therefore, we chose not to include them in our evaluation.

\noindent \textbf{Graph manipulation attack.} 
We provide a detailed description of the experimental setup used to evaluate the robustness of MirGuard against graph manipulation attacks, which aim to evade detection by modifying either the graph structure or node features. We broadly categorize these attacks into two types: data pollution attacks that occur during the detection phase, and data poisoning attacks that take place during the model training phase. Following prior work \cite{han2021sigl,jia2024magic,cheng2024kairos,rehman2024flash}, we adopt five different attack scenarios for comprehensive evaluation.

\textit{(1) Data Pollution Attacks.} Data pollution attacks aim to manipulate graph structures during the detection phase to hide malicious behaviors. For this category, we implemented three types of graph manipulation attacks:
\begin{itemize}
       \item \textbf{Graph Feature Pollution Attack (GFPA).} Alters the features of malicious nodes to mimic those of benign nodes, thereby hiding malicious behavior and evading detection.
    \item \textbf{Graph Structure Pollution Attack (GSPA).} Selectively adds new edges between malicious nodes and benign nodes, thereby altering the graph structure to make malicious nodes appear similar to benign nodes.
    \item \textbf{Combined Graph Pollution Attack (CGPA).} Combines both malicious feature manipulation and malicious structure manipulation methods, simultaneously altering the features and structure of malicious nodes to maximize the concealment of malicious behavior and evade detection.
\end{itemize}

These attacks were simulated by perturbing malicious nodes and their surrounding structures within the victim graph, mimicking realistic attacker behavior aimed at tampering with the graph.

\textit{(2) Data Poisoning Attacks.} Data poisoning attacks target the training phase, where the attacker perturbs the graph data used for training to compromise the model's robustness. Considering the practical difficulty for attackers to access the model directly, we focused on two types of poisoning attacks:

\begin{itemize}
    \item \textbf{Structure Poisoning Attack (SPA)}: Perturbs a certain proportion of nodes and edges in the training graph by adding or modifying connections, thereby disrupting the original structural features.
    \item \textbf{Feature Poisoning Attack (FPA)}: Alters a certain proportion of node features in the training graph by swapping initial features between nodes, disrupting the feature distribution, and misleading the model during training.
\end{itemize}

In summary, we provide detailed information about the attacks in Table \ref{tab:graph_attacks_detail}, including the attack targets (nodes or edges) and the perturbation rate (\(y\)). Specifically, the perturbation rate 
(\(y\)) represents the proportion of nodes or edges manipulated within the entire graph structure. these attacks are constructed at the node level, while for graph-level detection since the Streamspot and Unicorn datasets only provide anomalous graphs rather than nodes, we extend the attacks to the graph level. Specifically, we randomly select nodes or edges in the graph to be detected for attacks based on the perturbation rate (\(y\)).

\noindent \textbf{Metrics.}
In evaluating the performance of MirGuard, we use a variety of common metrics to comprehensively assess the model's behavior under different tasks and experimental setups. The basic evaluation metrics include Recall (Rec), Precision (Pre), AUC (Area Under the Curve), F1 Score (F1), and Accuracy (Acc). Additionally, we introduced an Absolute Change Rate (ACR) as an extra metric to evaluate robustness, which is utilized in Figure \ref{fig:ablation_study}. These metrics provide a holistic understanding of the model's performance, covering its detection capability, classification effectiveness, and balance between different classes.

\subsection{MirGuard's Effectiveness (RQ1)} \label{eva: effectiveness}
In this section, to evaluate the detection performance of MirGuard and its baseline models, we use precision, F1-score, and recall as evaluation metrics. The experiments are conducted on the Unicorn and Streamspot datasets for graph-level detection and the DARPA dataset for node-level detection. MirGuard adopts a self-supervised training approach, where the model is trained on benign data and evaluated on malicious data for detection.

\begin{figure}[t]
  \centering
  \subfloat[ROC curves (Node).]{
      \includegraphics[width=0.48\linewidth]{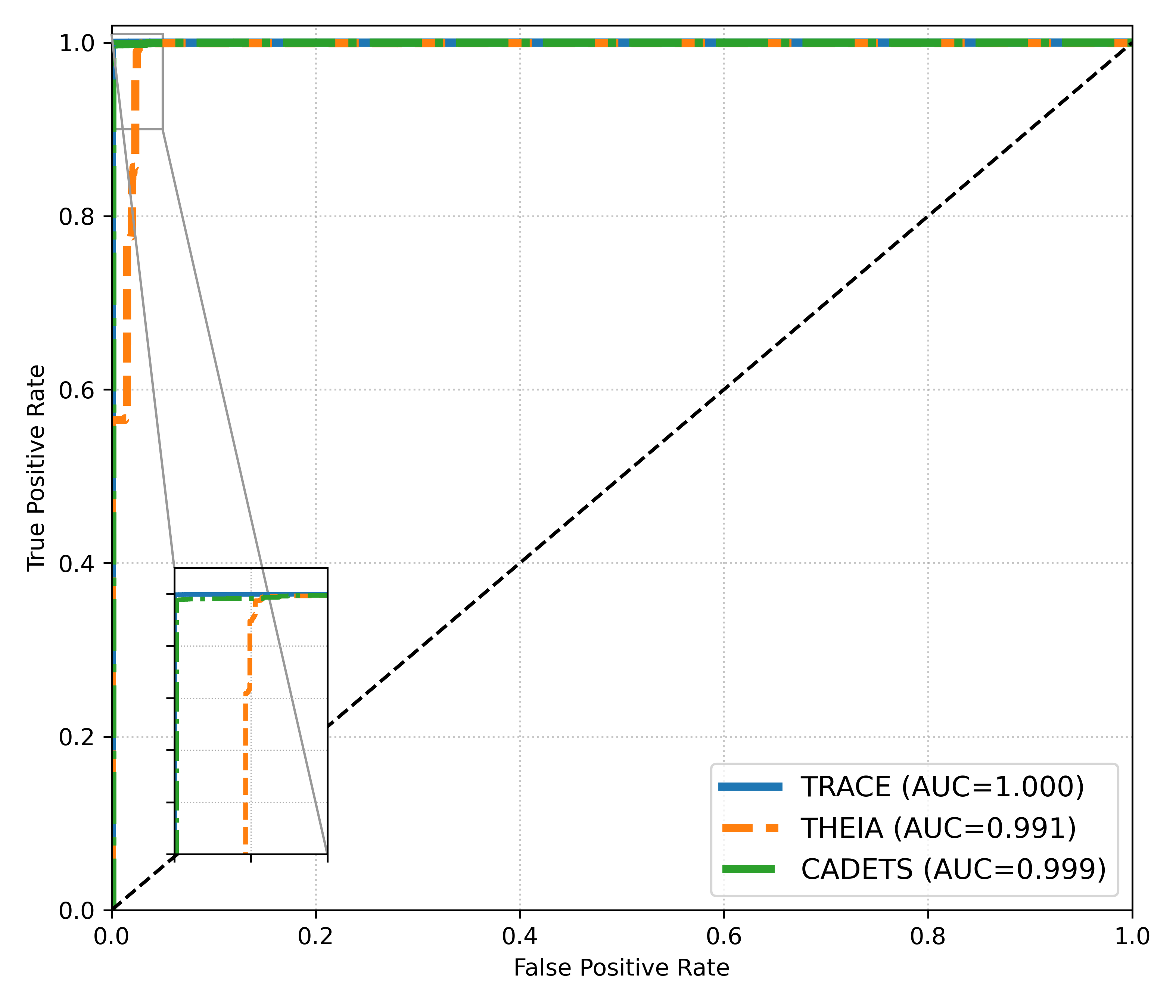} 
  }
    \subfloat[ROC curves (Graph).]{
      \includegraphics[width=0.48\linewidth]{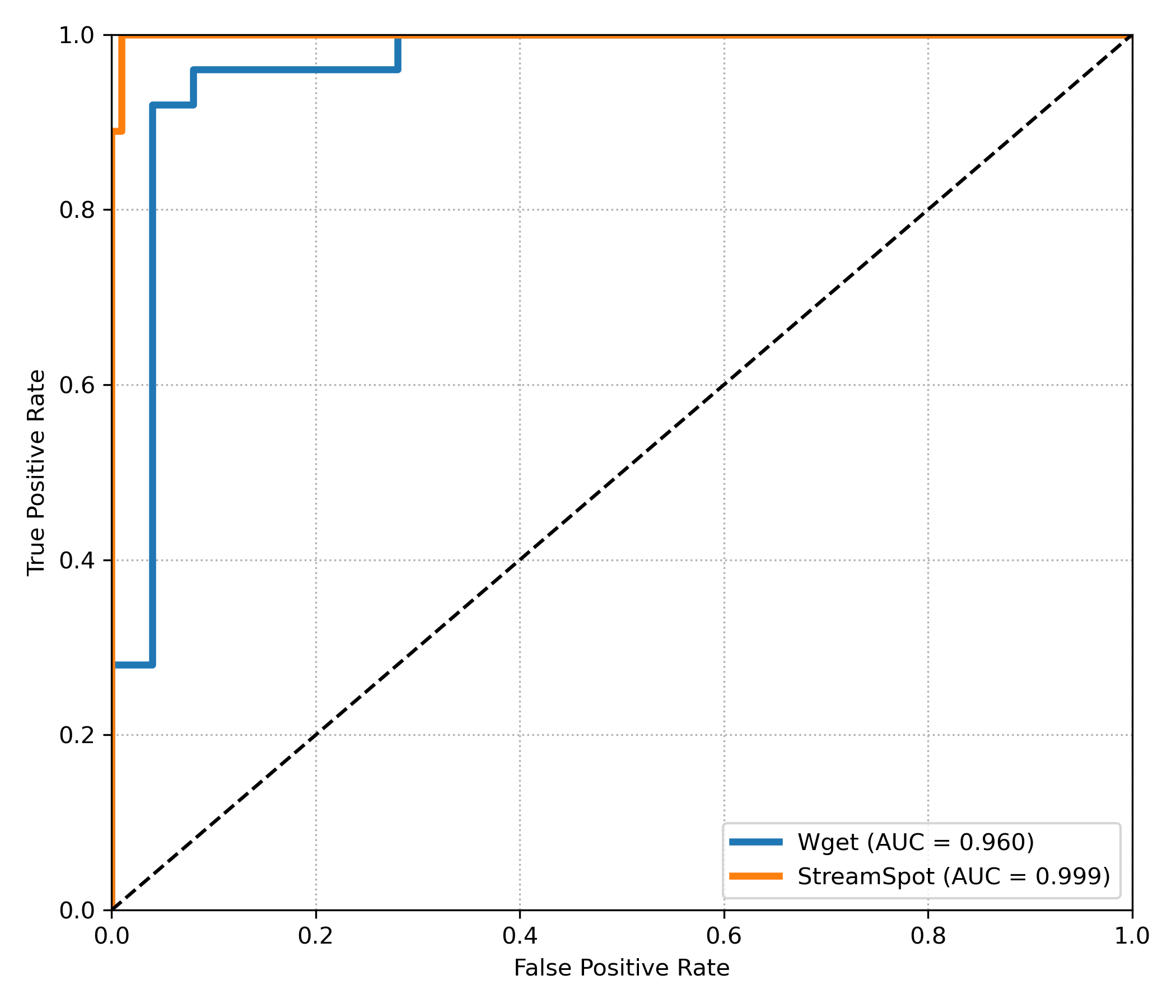} 
  }
  \caption{ROC curves on each dataset.}
  \label{fig:ROC}
\end{figure}

\noindent \textbf{Detection result.} Table \ref{tab:compare} provides the detection results of MirGuard, while Figure \ref{fig:ROC} shows the ROC curves for each dataset. In the graph-level anomaly detection datasets, Streamspot and Unicorn wget, MirGuard achieved near-perfect detection performance on the simpler Streamspot dataset, with a precision of 99\% and a recall of 100\%. This high performance is attributed to the dataset's collection of single-user activities per log batch, which are structurally and semantically distinct from each other.

On the more complex Unicorn Wget dataset, MirGuard still achieved high accuracy (96\%) and recall (96\%). Moving to node-level detection, MirGuard also demonstrated high performance on the DARPA TC datasets, achieving 99\% accuracy and 99\% recall. Due to the significant disparity between benign and malicious entities, MirGuard was able to accurately identify anomalies. This success is attributed to the use of KMeans for outlier detection, which effectively leverages the distinct feature distributions of benign and malicious entities.

\noindent \textbf{Comparison study.}
To compare the performance of MirGuard with existing state-of-the-art methods, as described in our experimental setup \ref{eva: setup}, we selected several detectors for both graph-level and node-level anomaly detection, including Threatrace \cite{wang2022threatrace}, FLASH \cite{rehman2024flash}, and MAGIC \cite{jia2024magic}.

As shown in Table~\ref{tab:compare}, MirGuard demonstrates outstanding performance across five representative provenance datasets. For node-level detection, it achieves near-perfect results on Theia, Cadets, and Trace. On Theia, MirGuard reaches 0.99 in both precision and F1-score, while achieving the lowest FPR of 0.03\%. On the Cadets and Trace datasets, it further reduces the FPR to below 0.01\%, outperforming both supervised detectors like Threatrace and FLASH, as well as advanced unsupervised methods like MAGIC. For graph-level detection, MirGuard also shows strong performance on both the Streamspot and Wget datasets. On Streamspot, it achieves a perfect recall of 1.00 and a balanced F1-score of 0.99, matching the results of MAGIC and FLASH. On the more complex Wget dataset, MirGuard leads all baselines with an FPR of only 0.6\%, significantly outperforming MAGIC and FLASH (2.0\%) and especially Threatrace (7.4\%). These results highlight the effectiveness of MirGuard’s contrastive learning framework, which leverages multi-view graph augmentation to generate robust and generalizable representations. Unlike MAGIC’s masked graph autoencoder approach that focuses on local reconstruction, MirGuard captures both local and global semantics by contrasting positive and negative views. This enables more reliable identification of stealthy or structurally evasive attack behaviors embedded in the provenance graph.

\begin{table}[t]
  \centering
  \caption{Comparison of Anomaly Detection Methods}
  \renewcommand{\arraystretch}{1.1} % 调整行距
  \resizebox{0.48\textwidth}{!}{
    \begin{tabular}{c|c|cccc}
    \hline
    \multirow{2}[2]{*}{Dataset} & \multirow{2}[2]{*}{Method} & \multicolumn{4}{c}{Metrics} \\
          &       & \multicolumn{1}{c}{Precision} & \multicolumn{1}{c}{F1-score} & \multicolumn{1}{c}{Recall} &  \multicolumn{1}{c}{FPR} \\
    \hline
    \multirow{4}[2]{*}{Theia} & Threatrace & 0.87  & 0.93  & \textbf{0.99} & 0.10\% \\
          & MAGIC & 0.98  &\textbf{ 0.99}  & \textbf{0.99} & 0.14\% \\
          & FLASH & 0.93  & 0.96  & \textbf{0.99} & 0.05\% \\
          & \textbf{MirGuard} & \textbf{0.99} & \textbf{0.99} & \textbf{0.99} & \textbf{0.03\%} \\
    \hline
    \multirow{4}[2]{*}{Cadets} & Threatrace & 0.90   & 0.95  & \textbf{0.99} & 0.20\% \\
          & MAGIC & 0.94  & 0.97  & \textbf{0.99} & 0.09\% \\
          & FLASH & 0.95  & 0.97  & \textbf{0.99} & 0.16\% \\
          & \textbf{MirGuard} & \textbf{0.98} & \textbf{0.99} & \textbf{0.99} &  \textless\textbf{0.01\%} \\
    \hline
    \multirow{4}[2]{*}{Trace} & Threatrace & 0.71  & 0.82  & \textbf{0.99} & 1.10\%\\
          & MAGIC & \textbf{0.99}  & \textbf{0.99}  & \textbf{0.99} & 0.09\%\\
          & FLASH & 0.95  & 0.97  & \textbf{0.99} & 0.16\%\\
          & \textbf{MirGuard} & \textbf{0.99} & \textbf{0.99} & \textbf{0.99} &  \textless\textbf{0.01\%} \\
    \hline
    \multirow{4}[2]{*}{Streamspot} & Threatrace & 0.98  & 0.99  & 0.99 & 0.4\% \\
          & MAGIC & 0.99  & 0.99  & \textbf{1.00} & 0.6\%\\
          & FLASH & \textbf{1.00}   & 0.96  & 0.98 & 0.3\% \\
          & \textbf{MirGuard} & 0.99 & \textbf{0.99} & \textbf{1.00} & 0.6\% \\
    \hline
    \multirow{4}[2]{*}{Wget} & Threatrace & 0.93  & 0.95  & 0.98 & 7.4\% \\
          & MAGIC & 0.96  & 0.95  & \textbf{0.96} & 2.0\%\\
          & FLASH & 0.96  & \textbf{0.96}  & \textbf{0.96} & 2.0\%  \\
          & \textbf{MirGuard} & \textbf{0.98} & \textbf{0.96} & \textbf{0.96} & \textbf{0.6\%}\\
    \hline
    \end{tabular}%
    }
  \label{tab:compare}
\end{table}%

\subsection{Adversarial Robustness Analysis (RQ2)} \label{eva: adver}

\begin{table*}[t]
  \centering
  \caption{Multi-various Graph Manipulation Attacks under Different Attack Rates.}
     \renewcommand{\arraystretch}{1.1} % 调整行距
    \begin{tabular}{c|c|ccc|ccc|ccc|ccc}
    \hline
     \multirow{2}[2]{*}{Attack} & \multirow{2}[2]{*}{Rate(\%)} & \multicolumn{3}{c|}{Threatrace \cite{wang2022threatrace}} & \multicolumn{3}{c|}{MAGIC \cite{jia2024magic}} & \multicolumn{3}{c|}{FLASH \cite{rehman2024flash}} & \multicolumn{3}{c}{MirGuard} \\
          &       & Precision & F1-score    & AUC   & Precision & F1-score  & AUC   & Precision & F1-score  & AUC   & Precision & F1-score & AUC \\
    \hline
    None  & \textbackslash{} & 0.904 & 0.949 & 0.954 & 0.944 & 0.970  & 0.997 & 0.947 & 0.972 & 0.978 & \textbf{0.981} & \textbf{0.989} & \textbf{0.999} \\
      \hline
    \multirow{3}[2]{*}{GSPA} & 10    & 0.731 & 0.813 & 0.910  & 0.734 & 0.848 & 0.921 & 0.817 & 0.837 & 0.921 & \textbf{0.978} & \textbf{0.942} & \textbf{0.996} \\
          & 20    & 0.617 & 0.749 & 0.854 & 0.644 & 0.800   & 0.862 & 0.723 & 0.759 & 0.889 & \textbf{0.957} & \textbf{0.932} & \textbf{0.984} \\
          & 50    & 0.307 & 0.489 & 0.756 & 0.334 & 0.533 & 0.745 & 0.593 & 0.657 & 0.828 & \textbf{0.861} & \textbf{0.887} & \textbf{0.972} \\
      \hline
    \multirow{3}[2]{*}{GFPA} & 10    & 0.784 & 0.878 & 0.913 & 0.904 & 0.959 & 0.991 & 0.887 & 0.922 & 0.952 & \textbf{0.979} & \textbf{0.988} & \textbf{0.999} \\
          & 20    & 0.744 & 0.843 & 0.907 & 0.873 & 0.950  & 0.979 & 0.840  & 0.896 & 0.941 & \textbf{0.976} & \textbf{0.976} & \textbf{0.998} \\
          & 50    & 0.644 & 0.797 & 0.871 & 0.794 & 0.920  & 0.957 & 0.793 & 0.871 & 0.938 & \textbf{0.967} & \textbf{0.963} & \textbf{0.988} \\
      \hline
    \multirow{3}[2]{*}{CGPA} & 10    & 0.767 & 0.845 & 0.901 & 0.784 & 0.870  & 0.974 & 0.807 & 0.877 & 0.934 & \textbf{0.971} & \textbf{0.931} & \textbf{0.997} \\
          & 20    & 0.693 & 0.749 & 0.882 & 0.713 & 0.830  & 0.913 & 0.787 & 0.841 & 0.913 & \textbf{0.953} & \textbf{0.937} & \textbf{0.989} \\
          & 50    & 0.484 & 0.489 & 0.824 & 0.527 & 0.655 & 0.819 & 0.667 & 0.777 & 0.865 & \textbf{0.873} & \textbf{0.872} & \textbf{0.975} \\
      \hline
    \multirow{3}[2]{*}{SPA} & 10    & 0.783 & 0.803 & 0.882 & 0.769 & 0.807 & 0.958 & 0.876 & 0.808 & 0.895 & \textbf{0.970} & \textbf{0.983} & \textbf{0.995} \\
          & 20    & 0.674 & 0.739 & 0.842 & 0.628 & 0.740  & 0.873 & 0.677 & 0.668 & 0.830  & \textbf{0.949} & \textbf{0.937} & \textbf{0.983} \\
          & 50    & 0.494 & 0.589 & 0.761 & 0.571 & 0.631 & 0.815 & 0.572 & 0.522 & 0.753 & \textbf{0.871} & \textbf{0.899} & \textbf{0.962} \\
    \hline
    \multirow{3}[2]{*}{FPA} & 10    & 0.834 & 0.821 & 0.904 & 0.904 & 0.960  & 0.983 & 0.877 & 0.952 & 0.953 & \textbf{0.980} & \textbf{0.989} & \textbf{0.999} \\
          & 20    & 0.785 & 0.777 & 0.895 & 0.884 & 0.940  & 0.853 & 0.853 & 0.946 & 0.948 & \textbf{0.978} & \textbf{0.970} & \textbf{0.998} \\
          & 50    & 0.744 & 0.759 & 0.874 & 0.807 & 0.890  & 0.916 & 0.790  & 0.931 & 0.922 & \textbf{0.933} & \textbf{0.951} & \textbf{0.981} \\
      \hline
    \end{tabular}%
  \label{tab:adv_ex}%
\end{table*}%

In this section, we conducted a comprehensive and in-depth evaluation of the robustness of MirGuard against graph tampering attacks. Table~\ref{tab:adv_ex} presents a comparative analysis of various attack types under different perturbation ratios. Threatrace and FLASH exhibit significant performance degradation under structural perturbations, especially at a 50\% attack ratio, where their F1 scores drop to 0.489 and 0.657, respectively. This indicates their strong reliance on local neighborhood structures, making them vulnerable to shifts in the embedding space caused by adversarial modifications. MAGIC, which adopts a node-masking strategy, shows improved local robustness and performs reasonably well under low-intensity attacks (0.8 F1 under GSPA=20\%). However, its performance deteriorates notably under structurally intensive global perturbations (i.e., higher attack ratios), revealing its lack of explicit global structural consistency enforcement.

In contrast, MirGuard, enhanced by multi-view perturbation strategies across nodes, edges, and features, consistently outperforms all baselines across all attack types and ratios. For instance, under CGPA-50\%, MirGuard achieves an F1 score of 0.872 and an AUC of 0.975, significantly surpassing other methods. Its strength lies in guiding the model to learn globally robust representations, thereby mitigating the impact of structural manipulations.

Furthermore, we evaluated the robustness of all models under increasing attack ratios (10\%, 20\%, 50\%). The results show that although all methods experience some performance drop under stronger perturbations, MirGuard exhibits the least degradation, consistently maintaining AUC \> 0.96 and F1 \> 0.87 in nearly all cases. It is also important to note that large-scale perturbations are often difficult to execute stealthily in real-world scenarios and tend to leave more forensic traces. Prior studies~\cite{Kunal2023Evading,sang2024obfuscating} have also indicated that high-ratio structural manipulations are challenging to realize in practice. Therefore, MirGuard demonstrates superior robustness against existing graph manipulation attacks.

% \begin{figure*}[t]
% \centering
% 	\subfloat[GSPA]{\includegraphics[width = 0.198\textwidth]{figure/GSPA.png}}
% 	\hfill
% 	\subfloat[GFPA]{\includegraphics[width = 0.198\textwidth]{figure/GFPA.png}}
%     \hfill
%     \subfloat[CGPA]{\includegraphics[width = 0.198\textwidth]{figure/CGPA.png}}
%     \hfill
%     \subfloat[SPA]{\includegraphics[width = 0.198\textwidth]{figure/SPA.png}}
%     \hfill
%     \subfloat[FPA]{\includegraphics[width = 0.198\textwidth]{figure/FPA.png}}
% \caption{Multi-various Graph Manipulation Attacks under Different Attack Rates (Cadets).}
% \label{fig: hyper-node-attack}
% \end{figure*}

% \begin{figure*}[t]
% \centering
% 	\subfloat[GSPA]{\includegraphics[width = 0.198\textwidth]{figure/GSPA_g.png}}
% 	\hfill
% 	\subfloat[GFPA]{\includegraphics[width = 0.198\textwidth]{figure/GFPA_g.png}}
%     \hfill
%     \subfloat[CGPA]{\includegraphics[width = 0.198\textwidth]{figure/CGPA_g.png}}
%     \hfill
%     \subfloat[SPA]{\includegraphics[width = 0.198\textwidth]{figure/SPA_g.png}}
%     \hfill
%     \subfloat[FPA]{\includegraphics[width = 0.198\textwidth]{figure/FPA_g.png}}
% \caption{Multi-various Graph Manipulation Attacks at Graph-Level under Different Attack Rates (Wget).}
% \label{fig: hyper-graph-attack}
% \end{figure*}

\begin{figure*}[t]
  \centering
  \tiny
  \subfloat[Wget, None]{
      \includegraphics[width=0.19\linewidth]{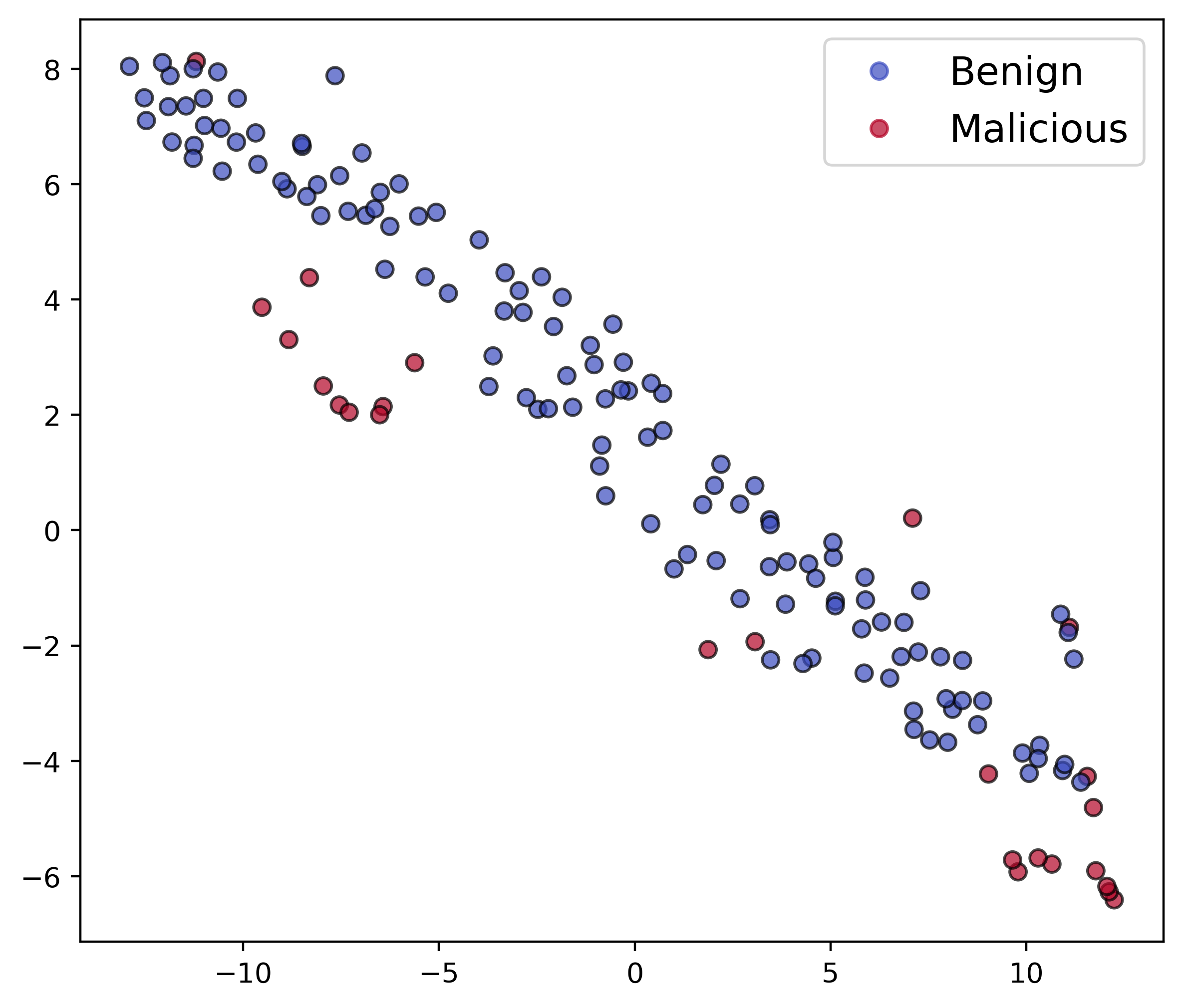} 
  }
  \subfloat[Streamspot, None]{
      \includegraphics[width=0.19\linewidth]{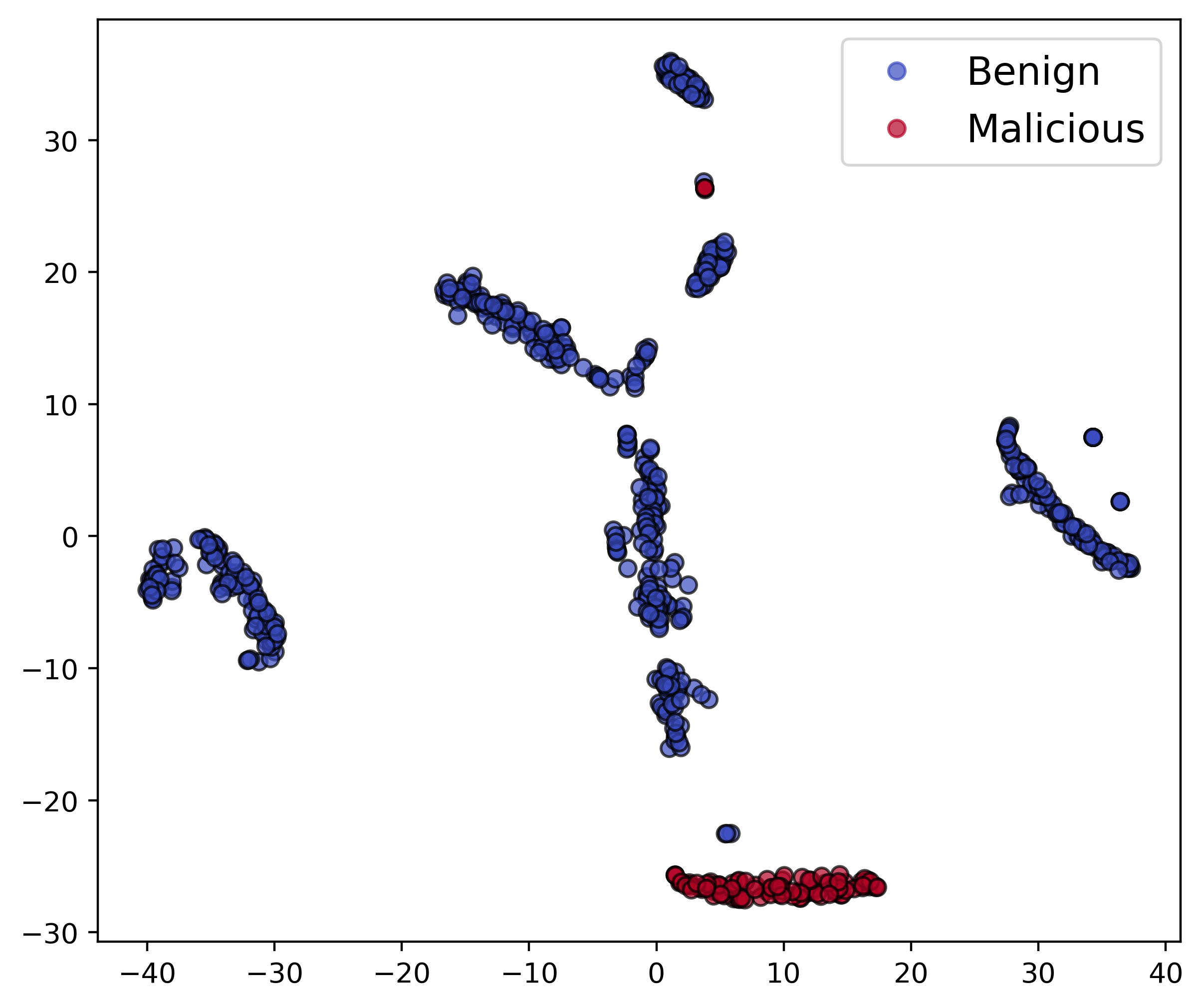} 
  }
    \subfloat[Cadets, None]{
      \includegraphics[width=0.19\linewidth]{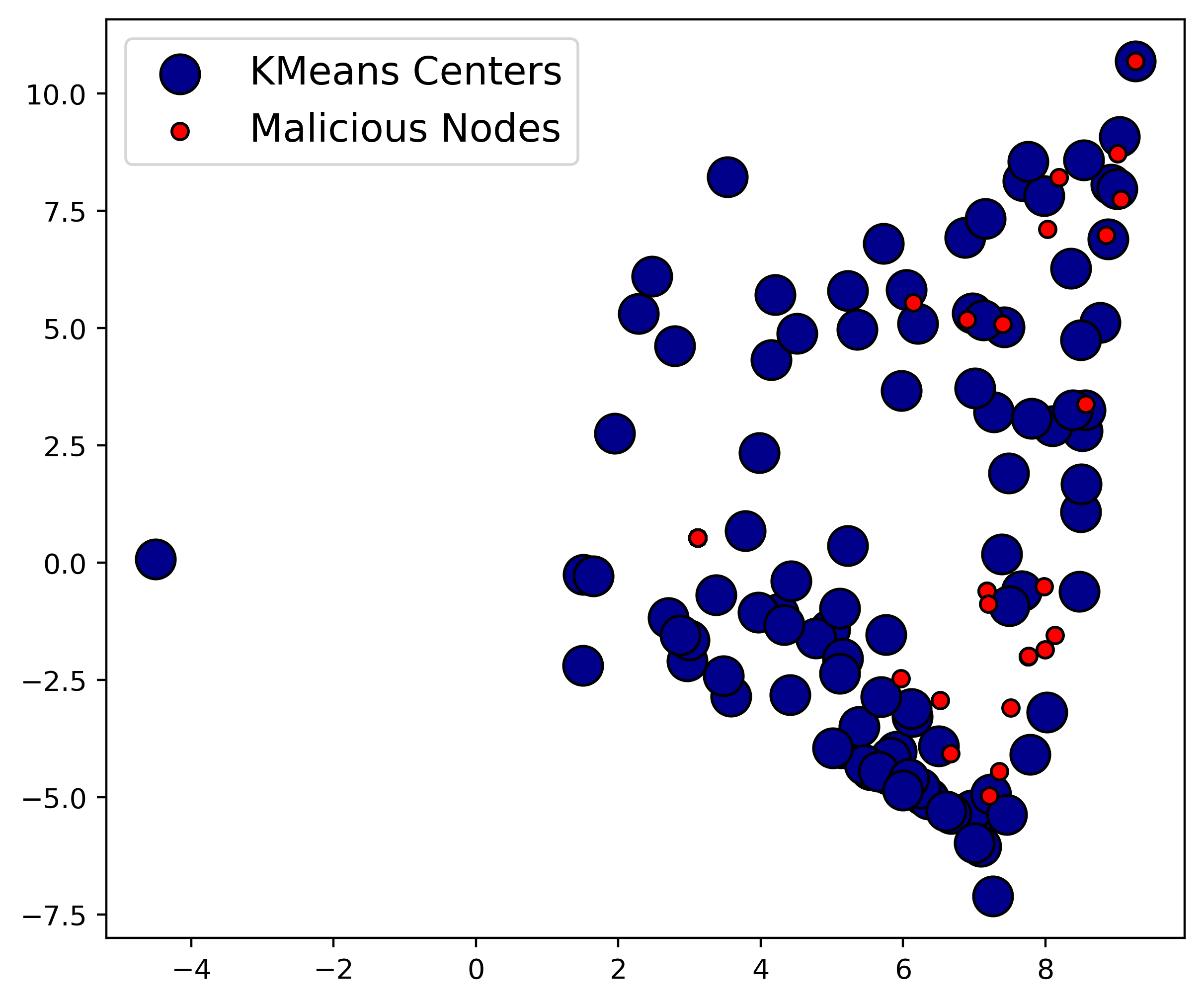} 
  }
    \subfloat[Trace, None]{
      \includegraphics[width=0.19\linewidth]{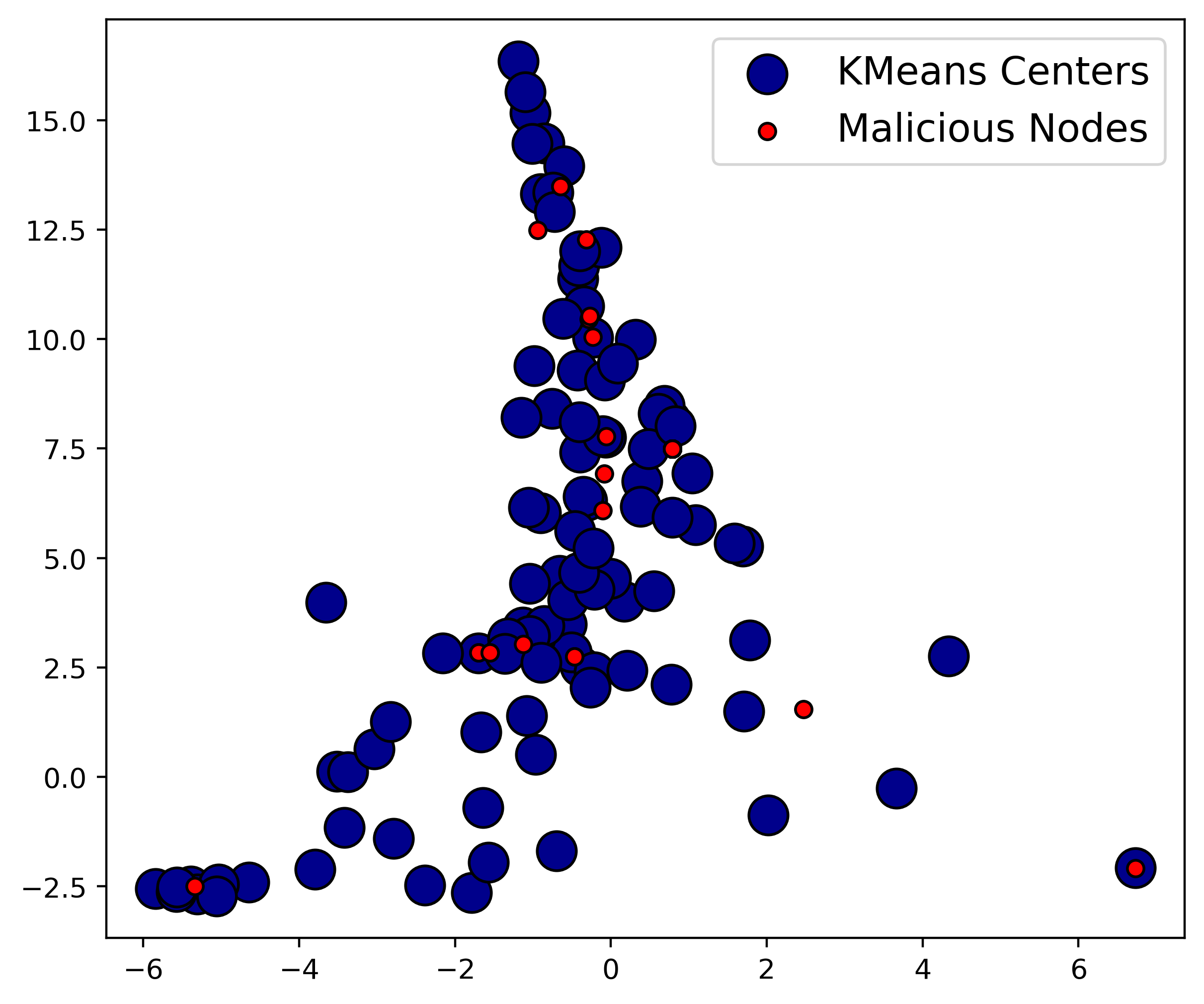} 
  }
  \subfloat[Theia, None]{
      \includegraphics[width=0.19\linewidth]{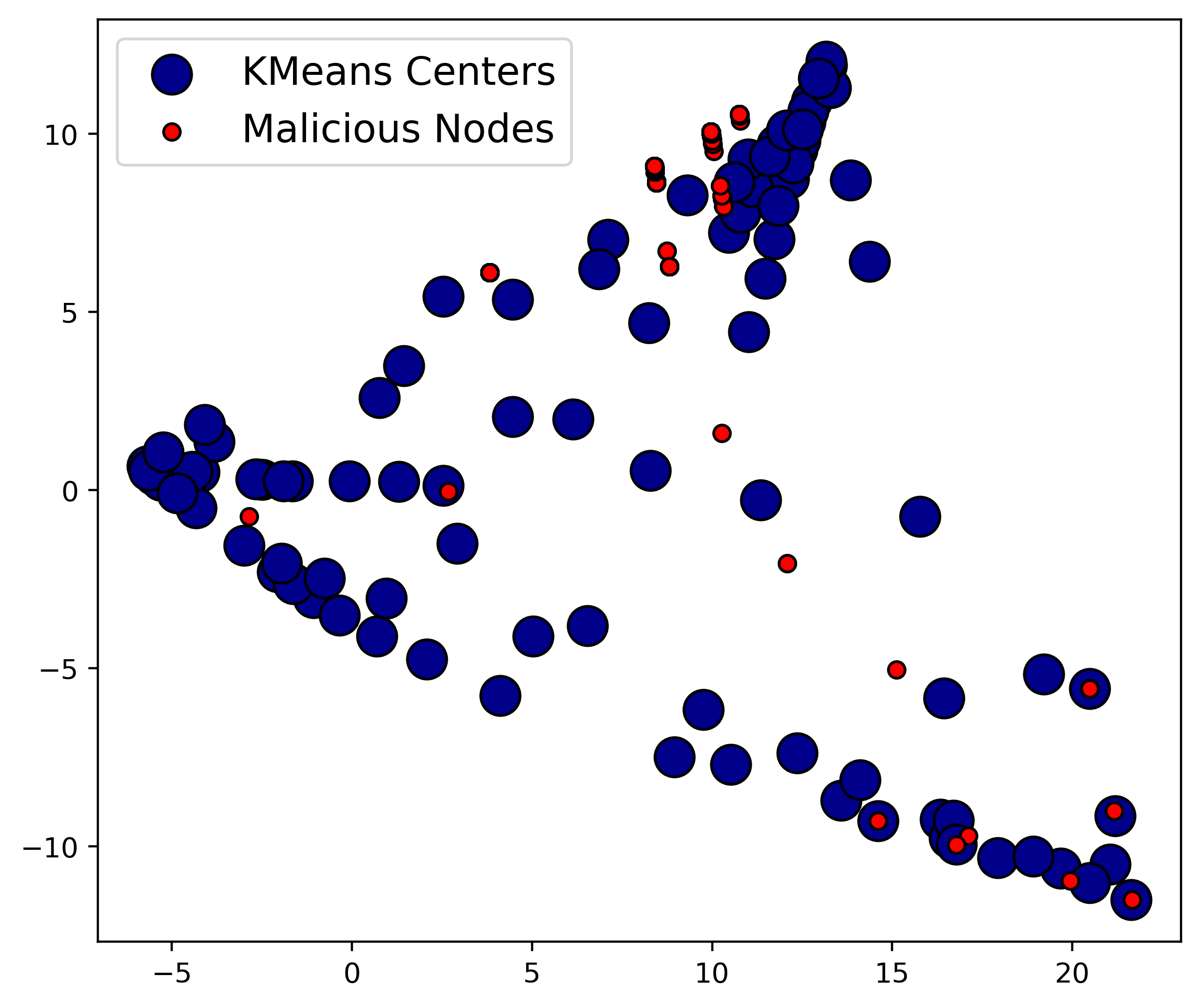} 
  }
  
    \centering
  \subfloat[Wget, \(y\)=(20\%)]{
      \includegraphics[width=0.19\linewidth]{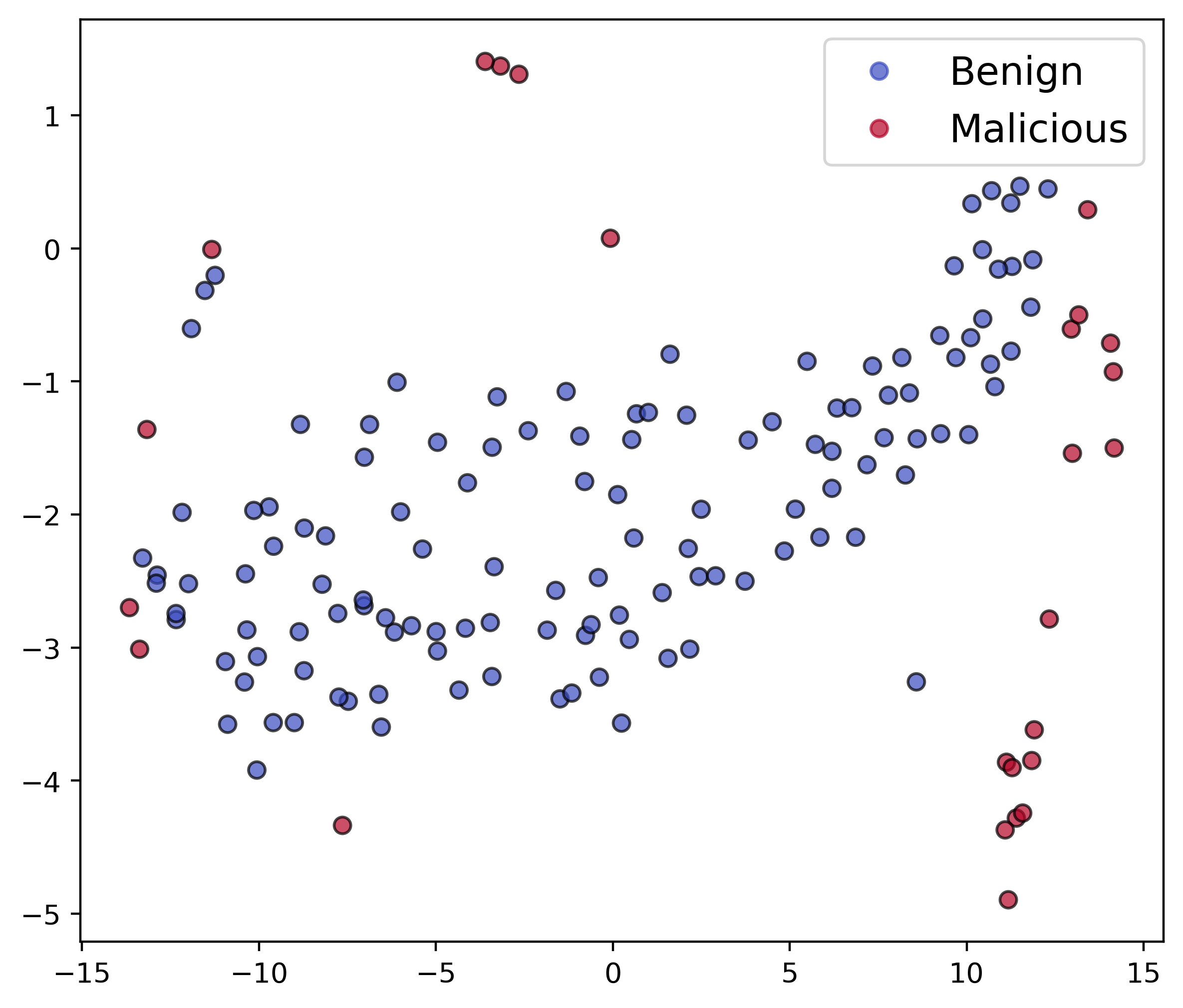} 
  }
  \subfloat[Streamspot,\(y\)=(20\%)]{
      \includegraphics[width=0.19\linewidth]{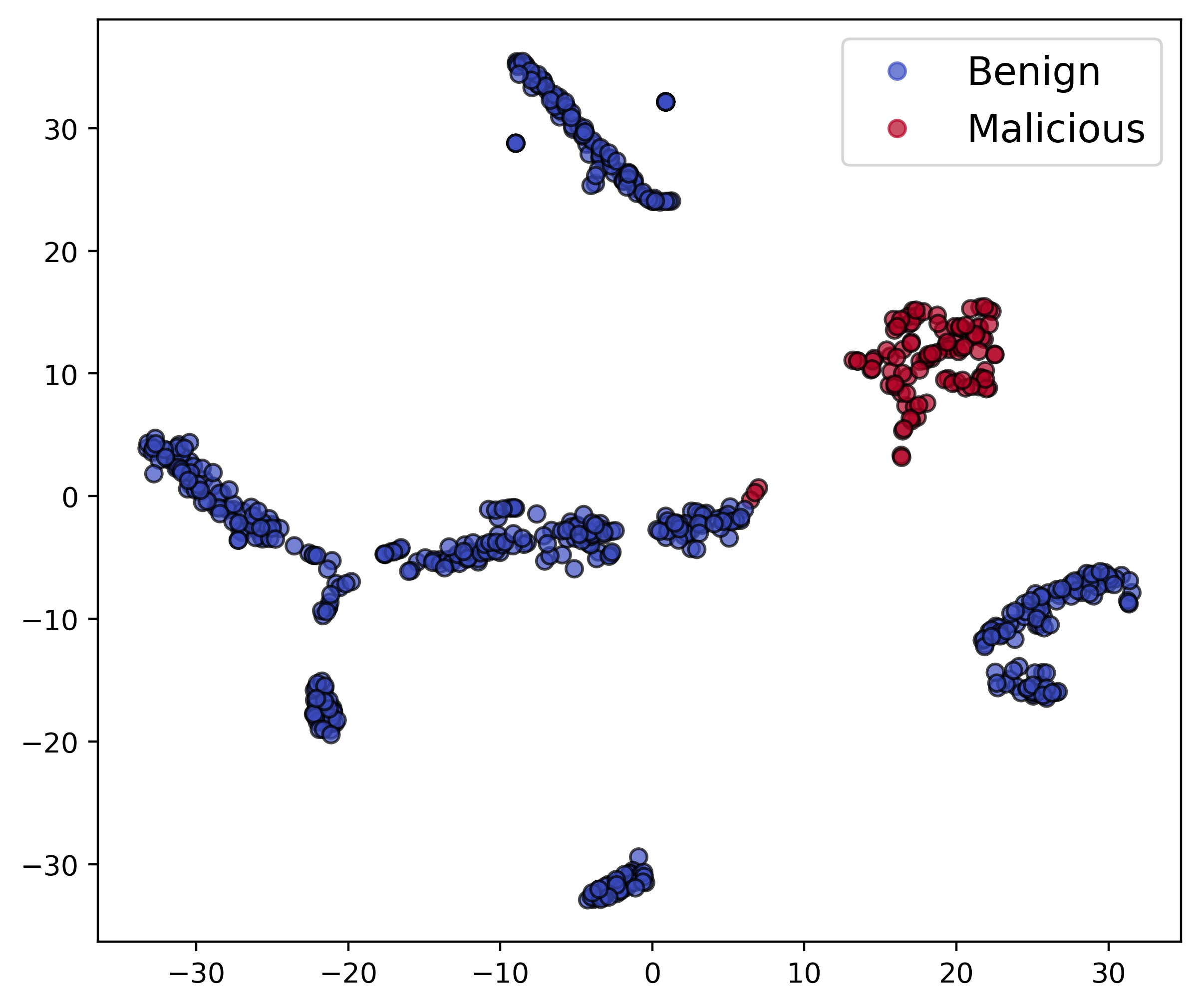} 
  }
   \subfloat[Cadets,\(y\)=(20\%)]{
      \includegraphics[width=0.19\linewidth]{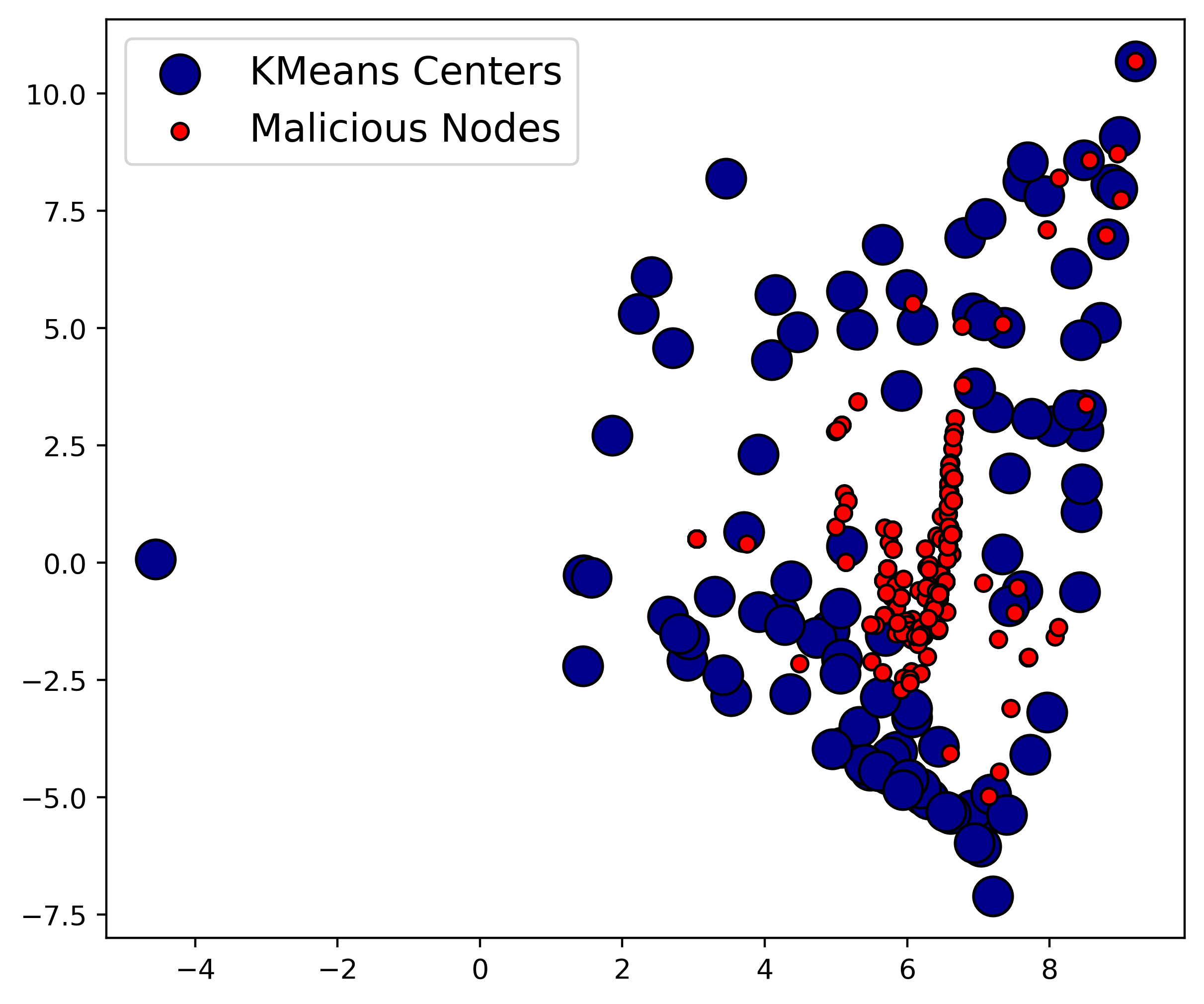} 
  }
    \subfloat[Trace,\(y\)=(20\%)]{
      \includegraphics[width=0.19\linewidth]{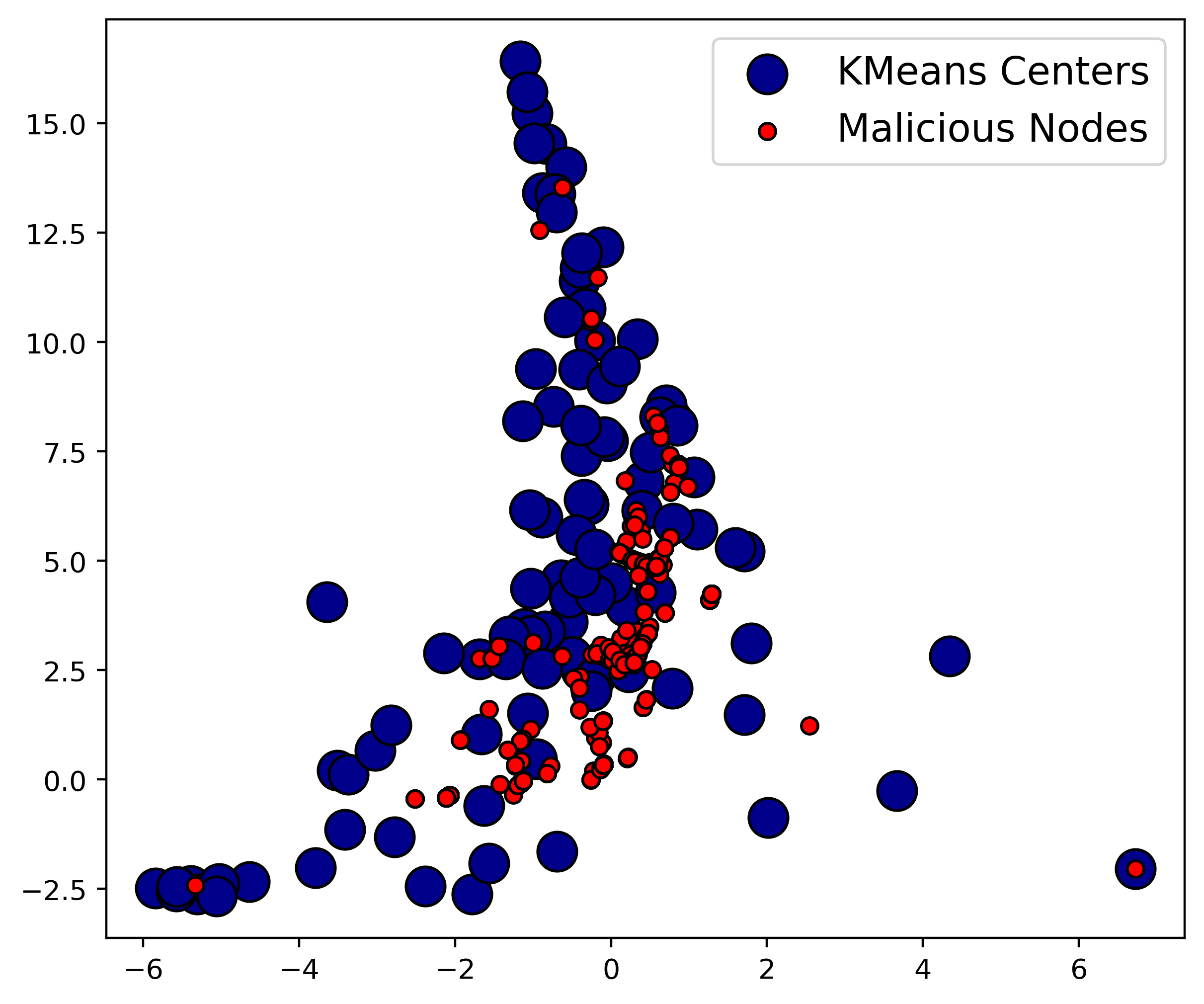} 
  }
   \subfloat[Theia,\(y\)=(20\%)]{
      \includegraphics[width=0.19\linewidth]{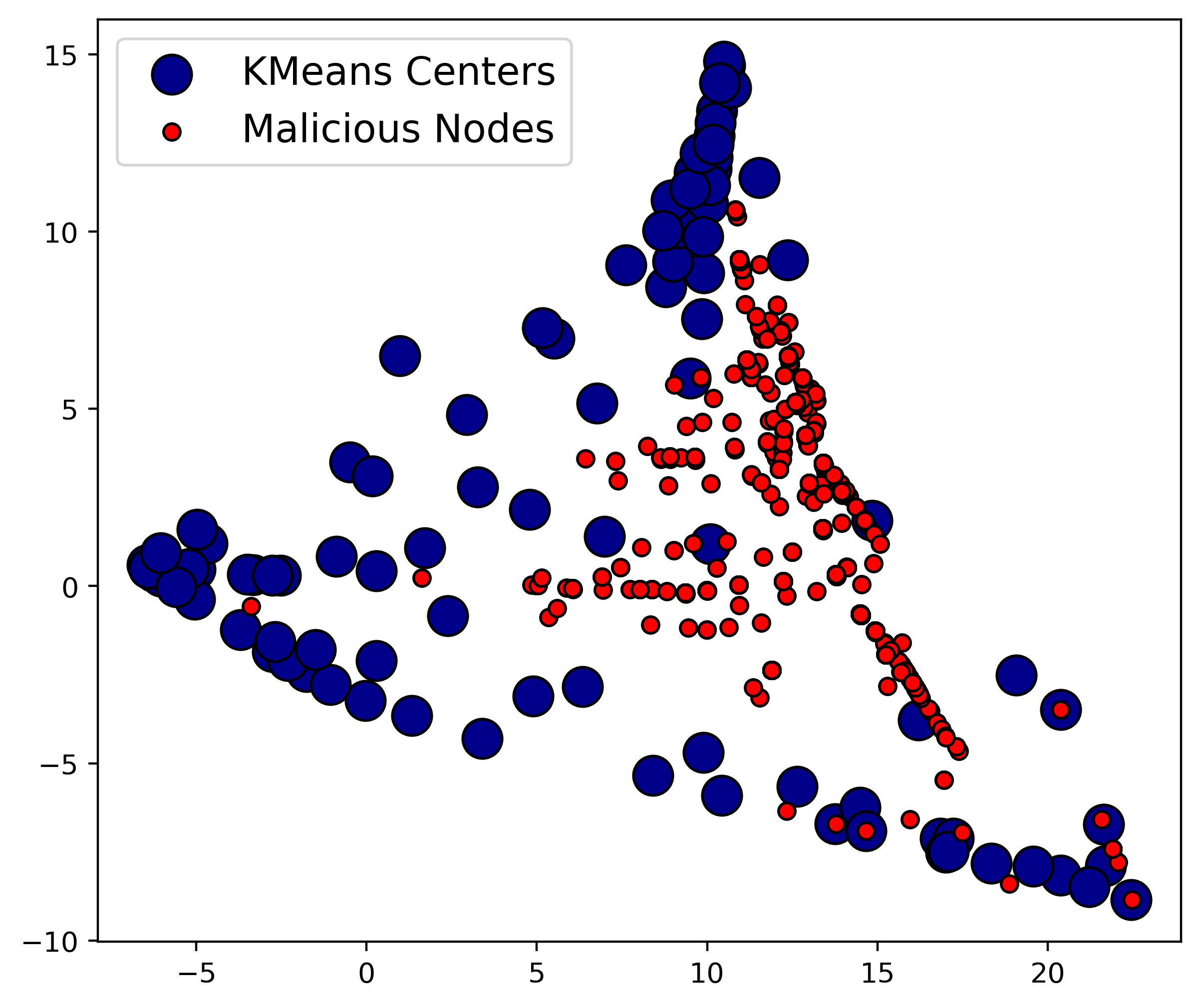} 
  }
  
  \caption{The latent representation learned by MirGuard has good discriminability and is able to resist graph manipulation attacks against the model.}
  \label{fig:tnse}
\end{figure*}

\noindent \textbf{Visualization of Representations.} The main contribution of MirGuard lies in its ability to learn high-quality representations that provide a comprehensive understanding of behavioral information, forming a clear decision boundary that effectively distinguishes between benign and malicious behavior nodes. To further analyze the internal representations learned by MirGuard, we visualized its latent representations under a graph manipulation attack (CGPA with an attack ratio of 0.2). We employed the t-SNE technique to project the graph representations of each input sample onto a 2D space. Figure \ref{fig:tnse} presents the learned representations, where subfigures (a), (b), (c), (d), and (e) depict the feature space distribution in the absence of attacks, while subfigures (f), (g), (h), (i), and (j) illustrate the feature space distribution after the attack.

In the figure, blue points represent benign samples, while red points represent malicious samples. Notably, in the DARPA dataset, due to the large number of nodes, we utilized K-means cluster centroids to approximate the distribution of benign nodes, while the red points indicate malicious nodes. Any node significantly distant from all cluster centroids is considered malicious. It can be observed that MirGuard’s learned latent representations are well-structured, allowing for a clear distinction between benign and malicious samples. Importantly, this compact representation and well-defined decision boundary not only enhance the model’s performance in classification tasks but also significantly increase the difficulty of adversarial attacks, thereby improving its robustness. After the graph manipulation attack, although some dispersion in the sample distribution is observed, MirGuard still effectively distinguishes between benign and malicious nodes. This can be attributed to MirGuard’s use of contrastive learning, which brings samples from different augmented views closer together, enabling the model to focus on global features rather than local features. As a result, the impact of graph manipulation attacks on MirGuard remains minimal.

\subsection{Ablation Study (RQ3)} \label{evl: ablation}
This section aims to investigate whether different modules in MirGuard, such as multi-view graph augmentation and contrastive learning, can improve its robustness and detection performance. Specifically, we evaluate the robustness of the current design by replacing certain components in MirGuard. Since graph augmentation and contrastive learning are designed to work collaboratively, we demonstrate the necessity of this design from two perspectives. First, we highlight the robustness of MirGuard's graph representation learning by substituting different methods in the representation learning module. Second, we verify the effectiveness of the graph augmentation strategy by introducing various augmentation techniques. Additionally, we further explore the impact of augmentation rate selection and the choice of detection methods.

\begin{table}[t]
  \centering
  \caption{Ablation study on graph representation model}
    \renewcommand{\arraystretch}{1.1} % 调整行距
   \resizebox{0.48\textwidth}{!}{
    \begin{tabular}{l|cccc|c}
    \hline
    \multicolumn{1}{c|}{\multirow{2}[1]{*}{Models}} & \multicolumn{4}{c|}{None}     & CGPA(y=20\%) \\
          & Precision & Recall & F1    & AUC   & F1 \\
    \hline
    MirGuard(DGI) & 0.9   & 0.8   & 0.91  & 0.9   & 0.678 \\
    MirGuard(GraphSAGE) & 0.92  & 0.96  & 0.92  & 0.94  & 0.642 \\
    MirGuard(MGAE) & 0.95  & 0.92  & 0.94  & 0.95  & 0.839 \\
    MirGuard & \textbf{0.98} & \textbf{0.99} & \textbf{0.99} & \textbf{0.99} & \textbf{0.96} \\
    \hline
    \end{tabular}%
    }
      \label{tab:ablation_GCL}
  \label{tab:addlabel}%
\end{table}%

\noindent \textbf{Effective of GCL Model.} 
To evaluate the robustness of the graph representation learning module in MirGuard, we conducted comparative experiments by replacing it with alternative methods, including GraphSAGE \cite{hamilton2017GraphSAGE}, DGI \cite{velickovic2019DGI}, and MGAE \cite{tan2022mgae}. The detection module still adopted the KNN-based strategy. In the experiments, we replaced the GCL module with these embedding methods to learn graph representations and performed anomaly detection. The training and testing datasets remained the same, and all experiments were conducted on the Cadets dataset with the attack ratio set to 0.2 to assess both robustness and detection performance. The results are shown in Table \ref{tab:ablation_GCL}. The GCL module achieves the best detection performance. Moreover, the contrastive learning component plays a critical role in enhancing the model's robustness, exhibiting only minimal performance degradation. In comparison, the alternative approaches, namely DGI, GraphSAGE, and MGAE, all result in noticeable performance drops. Among them, MGAE shows the second-best robustness. A possible explanation is that MGAE utilizes a masked reconstruction mechanism that reconstructs the graph structure based on unmasked nodes. This allows some attacked nodes to be masked during training, thereby improving its resistance to certain types of graph manipulation attacks.

\begin{table}[t]
  \centering
  \renewcommand{\arraystretch}{1.1} % 调整行距
  \caption{F1 for Different Augmentation Methods under Various Attack Types(\(\gamma_{\text{EA}}, \gamma_{\text{NA}}, \gamma_{\text{FA}}\))}
    \begin{tabular}{c|cccc}   
    \hline
    \textbf{Augmentation} & AUC & AUC & AUC & AUC \\
    \textbf{Type} & (CGPA) & (GSPA) & (GFPA) & (None) \\
    \hline
    NA & 0.954 & \textbf{0.971} & 0.967 & 0.989 \\
    EA & 0.972 & 0.961 & 0.964 & 0.985 \\
    FA & 0.965 & 0.963 & 0.947 & 0.973 \\
   \hline
    NA+FA & 0.979 & \textbf{0.971} & \textbf{0.988} & 0.991\\
    EA+FA & \textbf{0.984} & 0.959 & 0.987 & 0.989 \\
    NA+EA+FA & \textbf{0.984} & \textbf{0.971} & 0.984 & \textbf{0.999} \\
    \hline
    \end{tabular}%
  \label{tab:adv_auc}%
\end{table}%

\noindent \textbf{Effective of Graph Augmentation.} 
MirGuard introduces three types of graph augmentation methods. These techniques, through contrastive learning, encourage the model to focus on global behaviors while ignoring local perturbations. To investigate the impact of these graph augmentation strategies on the robustness of MirGuard, we evaluated their effectiveness under a graph contrastive learning framework against different attacks, including GSPA, GFPA, and CGPA. Specifically, we trained the model using various augmentation strategies on the Cadets dataset and assessed its robustness. The experimental results are shown in Table \ref{tab:adv_auc}.

The results show that in the absence of attacks, applying all three augmentation methods achieves the best performance. Under attack scenarios, different augmentation strategies contribute differently to robustness. For example, NA achieves the highest AUC of 0.971 under GSPA attacks, while the combination of NA and FA performs best under GFPA attacks, achieving an AUC of 0.988. The combination of NA, EA, and FA demonstrates consistent robustness across all attack types, with AUC values ranging from 0.971 to 0.984. Overall, Table \ref{tab:adv_auc} indicates that well-designed augmentation strategies can effectively enhance MirGuard’s robustness against various attacks. It is worth noting that although the NA+EA+FA combination does not always yield the best result, possibly due to excessive perturbations interfering with feature extraction, selecting appropriate augmentation strategies can still significantly improve adversarial robustness.

\begin{figure}[tbp]
    \centering

    \subfloat[Ablation on $\gamma$]{
        \includegraphics[width=0.47\linewidth]{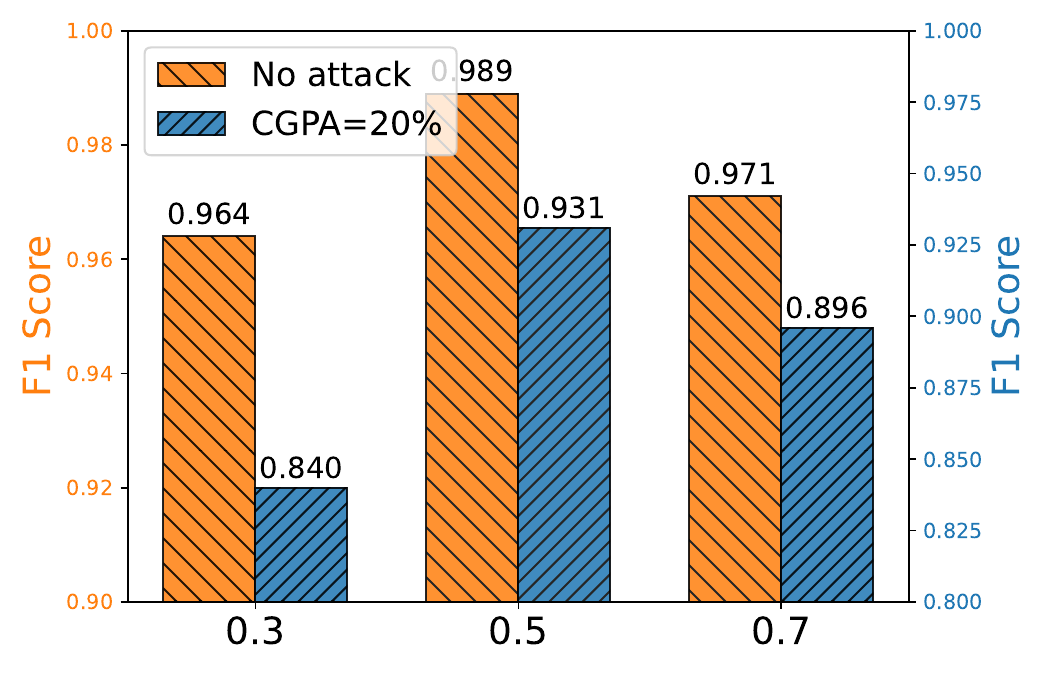} 
        \label{fig:ablation_modules}
    }
    \subfloat[Ablation on detector]{
        \includegraphics[width=0.47\linewidth]{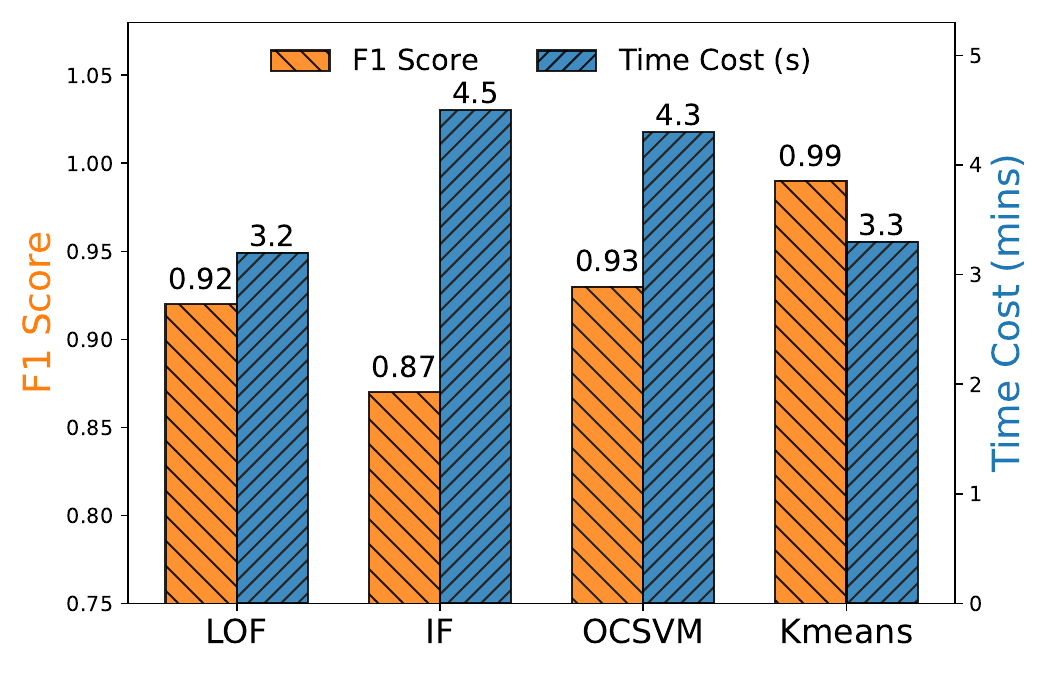} 
        \label{fig:ablation_gamma}
    }
    \caption{Ablation study on detector modules and augment rate $\gamma$ for MirGuard.}
    \label{fig:ablation_study}
\end{figure}

In addition, we explored the impact of the augmentation ratio on model robustness. We conducted experiments under both attack and non-attack scenarios. As shown in Figure \ref{fig:ablation_study}(a), as the augmentation ratio increases, the model’s performance first improves and then declines, and a similar trend is observed in its robustness. This is because a low perturbation ratio fails to effectively enhance the model’s learning capacity, while an excessively high ratio disrupts the original graph structure. Therefore, we recommend an optimal augmentation ratio of $\gamma = 0.5$.

\noindent \textbf{Effective of Detector.} We explored the classification performance of MirGuard by employing different classifiers. To this end, we configured several lightweight classifiers, including Local Outlier Factor (LOF) \cite{breunig2000lof}, One-Class Support Vector Machine (OCSVM) \cite{scholkopf1999OC-SVM}, and Isolation Forest (IF) \cite{liu2008isolation}. We conducted experiments with multiple detectors on the Cadets dataset, and the results are shown in Figure \ref{fig:ablation_study}.(b). The experimental results show that IF and OCSVM incurred relatively high time overhead and delivered moderate performance. In contrast, LOF had a similar time cost to KMeans but exhibited slightly lower accuracy. Overall, KMeans achieved the best classification performance while maintaining low time overhead.

\subsection{Performance Overhead (RQ4)} \label{evl: overhead}
Besides the robustness and effectiveness of MirGuard, efficiency is another critical factor influencing its practical application. In this section, we compare the training and inference costs of MirGuard with its baseline models to evaluate its efficiency. It is important to emphasize that during the training of these detectors, we used the default settings provided in their open-source implementations to achieve optimal performance, and the training process was conducted on the same server and configuration.

Table \ref{tab: overhead} summarizes the training costs of MirGuard and its baseline models. We observed that when trained with the same batch size, MirGuard and MAGIC exhibit similar time and memory overheads. On the other hand, FLASH and Threatrace adopt strategies that allow them to converge with smaller batch sizes and fixed graph sizes within each batch, which reduces memory overhead but results in longer training times. Regarding inference overhead, MirGuard maintains the best inference time and relatively low memory consumption. Therefore, overall, compared to the training costs of baseline detectors, it can be concluded that MirGuard ensures robustness without sacrificing efficiency.

\begin{table}[t]
  \centering
    \renewcommand{\arraystretch}{1.1} % 调整行距
  \caption{Performance Comparison of Different Methods in Terms of Training Time and Memory Usage}
  \resizebox{0.48\textwidth}{!}{%
    \begin{tabular}{l|l|cccc}
      \hline
      \textbf{Phase} &\textbf{Metric} & \textbf{FLASH} & \textbf{MAGIC} & \textbf{Threatrace} & \textbf{MirGuard} \\
      \hline
      
       \multirow{2}{*}{\textbf{Train}} & Total Time (s) & 4,580 & \textbf{151} & 2,780 & 214 \\
       & Memory (MB) & 760 & 1,564 & 1,031 & \textbf{1,525} \\
       \hline
      \multirow{2}{*}{\textbf{Inference}}  & Total Times (s) & 4,304 & 1,037 & 1,380 & \textbf{437} \\
       & Memory (MB) & \textbf{1,097} & 1,667 & 2,301 & 1,532 \\
      \hline
    \end{tabular}%
  }
  \label{tab: overhead}
\end{table}

\section{Related Work}

\noindent\textbf{Provenance-based IDS.} Recently, provenance-based IDS methods have been categorized into three main types: learning-based, statistical-based, and rule-based approaches. Statistical methods \cite{hassan2019nodoze,pasquier2017practical,fang2022back,liu2022rapid} model the anomaly degree of nodes using features such as temporal correlation, degree distribution, and rarity. Rule-based methods \cite{hossain2017sleuth,milajerdi2019holmes,hassan2020tactical,milajerdi2019poirot,zhu2023aptshield,hossain2020combating} create rules based on external knowledge to progressively match patterns in the provenance graph for anomaly detection. Learning-based approaches \cite{shen2019attack2vec,alsaheel2021atlas,liu2019log2vec,han2021sigl,yang2023prographer,zengy2022shadewatcher,wang2022threatrace,jia2024magic,cheng2024kairos} include sequence learning to extract and model sequence features for anomaly detection, as well as deep graph learning techniques for graph-level and edge-level detection using features like graph snapshots \cite{han2020unicorn,yang2023prographer,cheng2024kairos} and node interactions \cite{zengy2022shadewatcher}. Recent research has also explored node-level detection within provenance graphs, laying the foundation for fine-grained anomaly analysis. Threatrace \cite{wang2022threatrace} uses GraphSAGE for node embedding and anomaly detection, while MAGIC \cite{jia2024magic} employs MGAE for unsupervised graph representation learning and KNN-based anomaly detection. FLASH \cite{rehman2024flash} combines a GNN with Word2Vec for feature extraction and designs a caching mechanism to support scalability real-time detection.

\textbf{Graph Manipulation Attacks for Provenance-based Detector.} Wagner et al. \cite{wagner2002mimicry} first introduced mimicry attacks in 2002, enabling attackers to evade IDS detection. Their theoretical framework laid the foundation for circumventing PIDS. Li et al. \cite{2020Mimic} questioned the robustness of provenance-based detectors, highlighting risks like dependency explosion and proposing mimicry-based circumvention methods. Goyal et al. \cite{goyal2023sometimes} demonstrated the first practical evasion attacks against P-IDS in 2023. Kunal et al. \cite{Kunal2023Evading} advanced this with the PROVNINJA framework, reducing new system events and expanding tolerable distribution differences. Sang et al. \cite{sang2024obfuscating} proposed an obfuscation attack strategy, introducing meta-behavior mapping for realistic evasion, noting that large-scale graph tampering is impractical for attackers.

\textbf{Robustness of Graph Neural Networks.}
The robustness of GNN has gained increasing attention in recent years due to its wide applications in critical domains such as social networks, recommendation systems, and cybersecurity. However, studies have shown that GNNs are vulnerable to adversarial attacks, which manipulate graph structures, node features, or both to degrade model performance. These attacks are generally categorized into evasion attacks and poisoning attacks. Evasion attacks target the inference phase by perturbing graph data to mislead model predictions \cite{dai2018adversarial}, while poisoning attacks modify training data to undermine model robustness before deployment \cite{sun2022adversarial, xu2019topology}.

To address these threats, researchers have proposed various defense mechanisms. Adversarial training is one of the most widely studied methods, which enhances model robustness by injecting adversarial perturbations during training \cite{dai2018adversarial}. Additionally, graph data preprocessing techniques, such as graph sanitization \cite{wu2019adversarial}, aim to filter out adversarial perturbations, while robust GNN architectures leverage mechanisms like attention mechanisms or spectral filtering to strengthen resistance against attacks \cite{jin2020graph, guo2024rethinking}. Recently, contrastive learning has emerged as a promising direction to improve GNN robustness by combining data augmentations and contrasting positive and negative samples, effectively learning more robust graph embeddings \cite{you2020graphCL}.

\section{Discussion}
\noindent\textbf{Graph manipulation attacks.} Recently, graph manipulation attacks have posed significant challenges to the performance of provenance-based detectors \cite{goyal2023sometimes, Kunal2023Evading,sang2024obfuscating}. Attackers maliciously alter graph structures, causing the graph encoding of malicious nodes to resemble that of normal nodes, leading to false negatives. MirGuard demonstrates notable advantages in addressing these challenges, primarily due to our innovative approach of introducing various types of perturbations during the training phase and generating embeddings using contrastive learning. This method takes into account the potential manipulations of the graph structure by attackers during training, rendering attempts to alter the structure and features of malicious nodes ineffective. Therefore, MirGuard suggests a promising approach for countering such graph manipulation attacks in the future.

\section{Conclusion}

In this study, we introduced MirGuard, a novel graph learning-based anomaly detection system designed to enhance the robustness of provenance-based intrusion detection. By integrating multi-view augmentations with contrastive learning, MirGuard effectively mitigates mimicry attacks that manipulate graph structures. Comprehensive evaluations on multiple datasets demonstrate that MirGuard outperforms state-of-the-art detectors in both robustness and detection accuracy (achieving an average F1-score of over 96\% with less than 10\% AUC drop under graph manipulation attacks), without compromising efficiency (with overhead comparable to existing detectors). Our work provides a robust solution to modern cybersecurity challenges, paving the way for more robust provenance-based intrusion detection systems.

\bibliography{ref}
\bibliographystyle{ieeetr}


% \end{IEEEbiography}



\vfill

\end{document}